\documentstyle[12pt]{article}
\def\hybrid{\topmargin 0pt      \oddsidemargin 0pt
        \headheight 0pt \headsep 0pt
        \textwidth 17.5cm
        \textheight 25cm
        \voffset=-1.7cm
        \hoffset=-0.4cm
        \marginparwidth 0.0in
        \parskip 5pt plus 1pt   \jot = 1.5ex}
\catcode`\@=11
\def\marginnote#1{}

\newcount\hour
\newcount\minute
\newtoks\amorpm
\hour=\time\divide\hour by60
\minute=\time{\multiply\hour by60 \global\advance\minute by-\hour}
\edef\standardtime{{\ifnum\hour<12 \global\amorpm={am}%
        \else\global\amorpm={pm}\advance\hour by-12 \fi
        \ifnum\hour=0 \hour=12 \fi
        \number\hour:\ifnum\minute<10 0\fi\number\minute\the\amorpm}}
\edef\militarytime{\number\hour:\ifnum\minute<10 0\fi\number\minute}

\def\draftlabel#1{{\@bsphack\if@filesw {\let\thepage\relax
   \xdef\@gtempa{\write\@auxout{\string
      \newlabel{#1}{{\@currentlabel}{\thepage}}}}}\@gtempa
   \if@nobreak \ifvmode\nobreak\fi\fi\fi\@esphack}
        \gdef\@eqnlabel{#1}}
\def\@eqnlabel{}
\def\@vacuum{}
\def\draftmarginnote#1{\marginpar{\raggedright\scriptsize\tt#1}}

\def\draft{\oddsidemargin -0.1truein
        \def\@oddfoot{\sl preliminary draft \hfil
        \rm\thepage\hfil\sl\today\quad\militarytime}
        \let\@evenfoot\@oddfoot \overfullrule 3pt
        \let\label=\draftlabel
        \let\marginnote=\draftmarginnote
   \def\@eqnnum{{\rm (\theequation)}\rlap{\kern\marginparsep\tt\@eqnlabel}%
\global\let\@eqnlabel\@vacuum}  }

\font\teneuf=eufm10  scaled  1\@ptsize00 
\font\seveneuf=eufm7 scaled  1\@ptsize00 
\font\fiveeuf=eufm5  scaled  1\@ptsize00 

\newfam\euffam
\textfont\euffam=\teneuf  \scriptfont\euffam=\seveneuf
  \scriptscriptfont\euffam=\fiveeuf

\def\hexnumber@#1{\ifnum#1<10 \number#1\else
 \ifnum#1=10 A\else\ifnum#1=11 B\else\ifnum#1=12 C\else
 \ifnum#1=13 D\else\ifnum#1=14 E\else\ifnum#1=15 F\fi\fi\fi\fi\fi\fi\fi}

\def\got{\ifmmode\let\next\got@\else
 \def\next{\errmessage{Use \string\got\space only in math mode}}\fi\next}
\def\got@#1{{\got@@{#1}}}
\def\got@@#1{\fam\euffam#1}

\newfont{\lgot}{eufm10 scaled 1440}%

\newfont{\Bbb}{msbm10 scaled 1\@ptsize00}

\newcommand{\ZZ}{\mbox{\Bbb Z}}
\newfont{\Bbbb}{msbm7 scaled 1\@ptsize00}
\newcommand{\z}{\mbox{\Bbbb Z}}

\font\sevenmsa=msam6 
\newfam\msafam
\textfont\msafam=\sevenmsa
\def\hexnumber@#1{\ifnum#1<10 \number#1\else
\ifnum#1=10 A\else\ifnum#1=11 B\else\ifnum#1=12 C\else
\ifnum#1=13 D\else\ifnum#1=14 E\else\ifnum#1=15 F\fi\fi\fi\fi\fi\fi\fi}
\def\msa@{\hexnumber@\msafam}
\mathchardef\blacktriangleright="3\msa@49
\mathchardef\blacktriangleleft="3\msa@4A

\newdimen\linethick  \linethick=0.4pt
\newdimen\hboxitspace    \hboxitspace=5pt
\newdimen\vboxitspace    \vboxitspace=5pt

\def\fr#1{%
\beq\new
\vcenter{
\hrule height\linethick
           \hbox{\vrule width\linethick
                 \kern\hboxitspace
                 \vbox{\kern\vboxitspace
                       \hbox{$\begin{array}{c}\displaystyle#1
          \end{array}$}%
                       \kern\vboxitspace}%
                 \kern\hboxitspace
                 \vrule width\linethick}%
           \hrule height\linethick}%
\eeq}

\newdimen\Squaresize \Squaresize=14pt
\newdimen\Thickness \Thickness=0.5pt

\def\Square#1{\hbox{\vrule width \Thickness
   \vbox to \Squaresize{\hrule height \Thickness\vss
      \hbox to \Squaresize{\hss#1\hss}
   \vss\hrule height\Thickness}
\unskip\vrule width \Thickness}
\kern-\Thickness}

\def\Vsquare#1{\vbox{\Square{$#1$}}\kern-\Thickness}

\def\numberbysection{\@addtoreset{equation}{section}
        \def\theequation{\thesection.\arabic{equation}}}
\numberbysection

\newcommand{\sect}[1]{\setcounter{equation}{0}\section{#1}}
\renewcommand{\theequation}{\thesection.\arabic{equation}}
\newcommand{\l@qq}[2]{\addvspace{2em}
 \hbox to\textwidth{\hspace{1em}\bf #1 \dotfill #2}}

\newcounter{app}

\def\app{\setcounter{equation}{0}
\def\theequation{\Alph{app}.\arabic{equation}}\par
   \addvspace{4ex}
   \@afterindentfalse
  \secdef\@app\@dapp}

\newcommand\@app{\@startsection {app}{1}{0ex}%
                                   {-3.5ex \@plus -1ex \@minus -.2ex}%
                                   {2.3ex \@plus.2ex}%
                                   {\normalfont\Large\bf}}
\def\@dapp#1{%
{\parindent \z@ \raggedright  \bf #1}\par\nobreak}
\def\l@app#1#2{\ifnum \c@tocdepth >\z@
    \addpenalty\@secpenalty
    \addvspace{1.0em \@plus\p@}%
    \setlength\@tempdima{8em}%
    \begingroup
      \parindent \z@ \rightskip \@pnumwidth
      \parfillskip -\@pnumwidth
      \leavevmode \bfseries
      \advance\leftskip\@tempdima
      \hskip -\leftskip
      #1\nobreak\hfil \nobreak\hb@xt@\@pnumwidth{\hss #2}\par
    \endgroup\fi}
\newcounter{sapp}[app]

\def\sapp{\def\theequation{\Alph{app}.\arabic{equation}}
\par
\@afterindentfalse
  \secdef\@sapp\@dsapp}
\newcommand{\@sapp}{\@startsection{sapp}{2}{\z@}%
                                     {-3.25ex\@plus -1ex \@minus -.2ex}%
                                     {1.5ex \@plus .2ex}%
                                     {\normalfont\large\bfseries}}

\def\@dsapp#1{%
{\parindent \z@ \raggedright  \bf #1
}\par\nobreak}
\newcommand{\l@sapp}{\@dottedtocline{2}{1.5em}{2.3em}}

\def\titlepage{\@restonecolfalse\if@twocolumn\@restonecoltrue\onecolumn
     \else \newpage \fi \thispagestyle{empty}\c@page\z@
        \def\thefootnote{\fnsymbol{footnote}} }

\def\endtitlepage{\if@restonecol\twocolumn \else  \fi
        \def\thefootnote{\arabic{footnote}}
        \setcounter{footnote}{0}}  
\relax

\hybrid

\makeatletter
\newdimen\normalarrayskip              
\newdimen\minarrayskip                 
\normalarrayskip\baselineskip
\minarrayskip\jot
\newif\ifold             \oldtrue            \def\new{\oldfalse}
\def\arraymode{\ifold\relax\else\displaystyle\fi} 
\def\eqnumphantom{\phantom{(\theequation)}}     
\def\@arrayskip{\ifold\baselineskip\z@\lineskip\z@
     \else
     \baselineskip\minarrayskip\lineskip1\baselineskip\fi}

\def\@arrayclassz{\ifcase \@lastchclass \@acolampacol \or
\@ampacol \or \or \or \@addamp \or
   \@acolampacol \or \@firstampfalse \@acol \fi
\edef\@preamble{\@preamble
  \ifcase \@chnum
     \hfil$\relax\arraymode\@sharp$\hfil
     \or $\relax\arraymode\@sharp$\hfil
     \or \hfil$\relax\arraymode\@sharp$\fi}}

\def\@array[#1]#2{\setbox\@arstrutbox=\hbox{\vrule
     height\arraystretch \ht\strutbox
     depth\arraystretch \dp\strutbox
     width\z@}\@mkpream{#2}\edef\@preamble{\halign \noexpand\@halignto
\bgroup \tabskip\z@ \@arstrut \@preamble \tabskip\z@ \cr}%
\let\@startpbox\@@startpbox \let\@endpbox\@@endpbox
  \if #1t\vtop \else \if#1b\vbox \else \vcenter \fi\fi
  \bgroup \let\par\relax
  \let\@sharp##\let\protect\relax
  \@arrayskip\@preamble}
\def\eqnarray{\stepcounter{equation}%
              \let\@currentlabel=\theequation
              \global\@eqnswtrue
              \global\@eqcnt\z@
              \tabskip\@centering
              \let\\=\@eqncr
              $$%
 \halign to \displaywidth\bgroup
    \eqnumphantom\@eqnsel\hskip\@centering
    $\displaystyle \tabskip\z@ {##}$%
    &\global\@eqcnt\@ne \hskip 2\arraycolsep
         $\displaystyle\arraymode{##}$\hfil
    &\global\@eqcnt\tw@ \hskip 2\arraycolsep
         $\displaystyle\tabskip\z@{##}$\hfil
         \tabskip\@centering
    &{##}\tabskip\z@\cr}
\makeatother

\newtheorem{lem}{Lemma}[section]

\def\bea{\begin{eqnarray}}
\def\eea{\end{eqnarray}}
\def\nn{\nonumber}
\def\beq{\begin{equation}}
\def\eeq{\end{equation}}
\def\be{\beq\new\begin{array}{c}}
\def\ee{\end{array}\eeq}
\def\stackreb#1#2{\mathrel{\mathop{#2}\limits_{#1}}}
\def\square{\hfill{\vrule height6pt width6pt            
depth1pt} \break \vspace{.01cm}}                        
\def\d{\partial}

\def\comp{ \raise1pt\hbox{$\mathsurround=0pt\,\scriptstyle \circ\,$}}

\def\btl{\raisebox{1pt}{$\,\blacktriangleleft\,$}}
\def\tl{\triangleleft}

\def\sem{\raise1pt\hbox{$\scriptscriptstyle >\!$}\:\!\!\tl}
\def\dr{\raise1pt\hbox{$\scriptscriptstyle >\!$}\!\!\btl}
\def\]{]\raise-2pt\hbox{$_\ast$}}
\def\st#1{#1\raise-2pt\hbox{$_\ast$}}

\def\op#1{\raise-6pt\hbox{$\stackrel{\displaystyle\oplus }{\scriptstyle
#1}$}\;}
\def\oop#1#2{\raise-6pt\hbox{$\stackrel{#2}{\stackrel{\displaystyle\oplus
}{\scriptstyle #1}}$\;}}

\def\nH{\,\ov{\!H\!}\:}
\def\nH{\,\ov{\!H\!}\:}

\def\nT{\,\ov{\!T}}

\def\al{\alpha}
\def\la{\lambda}

\def\e{\epsilon}
\def\<{\langle}
\def\>{\rangle}
\def\ov{\overline}
\def\wt{\widetilde}

\def\cm{{\cal M}}
\def\cl{{\cal L}}
\def\cp{{\cal P}}
\def\cq{{\cal Q}}

\def\1{1\!\!1}
\def\Tr{\mbox{\footnotesize Tr}}

\def\L{\mbox{\bf L}}
\def\vir{\mbox{\footnotesize\bf L}}
\begin{document}
\begin{titlepage}
\thispagestyle{empty}
\hfill ITEP-TH-78/97 \\
\phantom. \hfill hepth/9810091 \\

\begin{center}
\vspace{0.1in}{\Large\bf Kadomtsev-Petviashvili Hierarchy and Generalized
Kontsevich Model}\\[.4in]
\bigskip {\large S.Kharchev}\\
\bigskip {\it Institute of Theoretical and Experimental
Physics,  \\
 Bol.Cheremushkinskaya st., 25, Moscow, 117 259},
 \footnote{E-mail address: kharchev@vitep5.itep.ru}
\end{center}
\bigskip
\bigskip
\begin{abstract}
The review is devoted to the integrable properties of the Generalized
Kontsevich Model which is supposed to be an universal matrix model to describe
the conformal field theories with $c<1$. The careful analysis of the model with
arbitrary polynomial potential of order $p+1$ is presented. In the case of
monomial potential the partition function is proved to be a $\tau$-function
of the $p$-reduced Kadomtsev-Petviashvili hierarchy satisfying $\L_{-p}$
Virasoro constraint. It is shown that the deformations of the "monomial"
phase to "polynomial" one have the natural interpretation in context of
so-called equivalent hierarchies. The dynamical transition between equivalent
integrable systems is exactly along the flows of the dispersionless Kadomtsev-
Petviashvili hierarchy; the coefficients of the potential are shown to be
directly related with the flat (quasiclassical) times arising in $N=2$
Landau-Ginzburg topological model. It is proved that the partition
function of a generic Generalized Kontsevich Model can be presented as a
product of "quasiclassical" factor and non-deformed partition function which
depends only on the sum of transformed integrable flows and flat times.
The Virasoro constraint for solution with an arbitrary potential is shown to be
a standard $\L_{-p}$-constraint of the (equivalent) $p$-reduced hierarchy
with the times additively corrected by the flat coordinates. The rich structure
of the model requires the implications almost all aspects of the classical
integrability. Therefore, the essential details of the fermionic approach to
Kadomtsev-Petviashvili hierarchy as well as the notions of the equivalent
integrable systems and their quasiclassical analogues are collected together
in parallel with step-by-step investigation of the suggested universal matrix
model.
\end{abstract}
\end{titlepage}
\newpage
\footnotesize
\tableofcontents
\newpage
\normalsize
\section{Introduction}
During the last years, the matrix models play an important role in the theory
of $2$-dimensional gravity, topological models and statistical physics
(see \cite{Mor} and references therein).
This paper is devoted to the study of particular 1-matrix model in an external
matrix field which is supposed to be "the universal" one .
The structure of the model is essentially defined by the matrix
integral of the typical form
\be\label{i1}
Z^V_{_N}[M]\,\sim\,\int dX\,e^{-\Tr\,V(X)+\Tr\,XV'(M)}
\ee
where $M$, $X$ are Hermitian $N\times N$ matrices and
$dX\sim\prod_{i,j=1}^NdX_{ij}$. In (\ref{i1})
$V(X)$ is an arbitrary potential (see exact formulation below).
The model with $V(X)=\frac{1}{3}X^3$ (the Kontsevich model) has been derived
in \cite{Kon} as a generating function of the intersection numbers on the
moduli spaces, i.e. by purely geometrical reasons, guided by Witten's
treatment of $2$-dimensional topological gravity \cite {Wit90}. Unfortunately,
the similar interpretation of more complicated model with an arbitrary
polynomial potential is still lacking.
Actually, the same model
(though in somewhat implicit form) have appeared for the first time in
\cite{KS} inspired by more "physical" arguments \cite{FKN1}, \cite{DVV}.
The advantage of the paper \cite{KS} consists of the fact that it starts
from the {\it integrable} properties of the model from the very beginning:
\cite{KS} gives the clear interpretation
of the Kontsevich partition function as a concrete solution of $2$-reduced
Kadomtsev-Petviashvili (KP) hierarchy, that is, the Korteveg-de-Vries one.
This allows to generalize the original Kontsevich model immediately.
In \cite{GKM} the partition function with an arbitrary potential has been
suggested as "the universal" matrix model under the name of Generalized
Kontsevich Model (GKM) (independently, the integral (\ref{i1}) with the
monomial potential of finite order has been considered in papers \cite{AM}-
\cite{IZ1}). The universality of GKM is based on the following facts
\cite{GKM},\cite{LGM}, \cite{versus}:
\begin{itemize}
\item[(i)] {For monomial potential it describes properly the
(sophisticated) double scaling limit of any multimatrix model.}
\item[(ii)] {The GKM partition function with a polynomial potential of order
$p+1$ is a $\tau$-function of the
Kadomtsev-Petviashvili hierarchy, properly reducible at the points, associated
with multi-matrix models, to a solution of $p$-reduced hierarchy. Moreover, it
satisfies an additional equation, which reduces
to conventional Virasoro constraint (string equation) for multi-matrix models
when potential
is degenerates to monomial one.}
\item[(iii)] {It allows the deformations of the potential
associated with a given multi-matrix model to potentials corresponding
to other models.}
\item[(iv)]
{GKM with arbitrary polynomial potential is directly connected with $N=2$
supersymmetric Landau-Ginzburg theories.}
\item[(v)] {The partition function
(\ref{i1}) with potential $V(X)\sim X^2+ n\log X$ describes the standard
1-matrix model before double-scaling limit.}
\item[(vi)] {Adding the negative
powers of $X$, the model gives a particular solution of Toda Lattice
(TL)hierarchy.}
\end{itemize}
Besides, the GKM is a non-trivial (and more or
less explicit) example of solution of the Kadomtsev-Petviashvili hierarchy
corresponding to the Riemann surface of the infinite genus. As an integrable
system with an infinite degrees of freedom, it possesses very rich structure
encoding many features which are absent for finite-dimensional systems: the
Virasoro and, more generally, $W$-constraints do not exhaust the complexity
of the model. It turns out that GKM is properly designed to describe the
quasiclassical (dispersionless) solutions parametrized by the coefficients
of the potential $V(X)$ being in the same time the exact solution of the
original hierarchy. This makes the study of GKM very promising from the
point of view of the physical applications as well as in the context of purely
mathematical aspects concerning the integrable structures.
In this paper we shall deal with the integrable properties of (\ref{i1})
only.

\bigskip\noindent
To investigate the GKM model in detail, the long way to overcome is required.
The point is that this model, being excellent example of the explicit
solution of the integrable system (KP or even TL hierarchy), unifies many
aspects of the latter. Besides well elaborated general strategy \cite{DJKM},
\cite{UT}, \cite{SP} to describe the above hierarchies, some more subtle
notions have to be implemented.
Originally author had a temptation to dump all the details concerning
the standard material to appendices (including the fermionic approach to
$\tau$-function). After some contemplation it become clear that in this
case the paper will contain the only Introduction as a main body with a lot
of appendices so the structure will be the same. Therefore, by the
pedagogical reasons, the decision arises to arrange the things as
self-consistent as possible.
The paper is organized as follows. In the first three sections we follow
the approach developed in \cite{DJKM}; the material here (except of
some details) is quite standard.
In Sect. 2 we give the essentials
concerning the most important integrable system, namely, the
Kadomtsev-Petviashvili hierarchy. We discuss briefly the
pseudo-differential calculus and introduce the notions of the Baker-Akhiezer
functions as well as the central object, the $\tau$-function, as a solution
of the evolution equations.
In Sect. 3 the fermionic approach to realization of $gl_\infty$ is given .
It is of importance while representing
the solutions of the KP hierarchy in the "explicit" form in terms
of the fermionic correlators. In Sect. 4 we represent the
$\tau$-function in the specific determinant form using the fermionic approach
introduced in the previous section. Such representation is very natural from
the Grassmannian approach to the integrable systems \cite{SP} and, what is more
important for the present purposes, this gives the simple proof of the
integrability of the GKM partition function.
The Generalized Kontsevich Model is introduced in Sect. 5. First of all,
simplify the GKM partition  function by the standard integration over the
angle variables thus obtaining
the integral over the eigenvalues $x_1,\,\ldots,\,x_{_N}$ of the matrix $X$.
After this, we are able to write the partition function in the
determinant form which is the starting point to investigate
its integrable properties. We prove that, in the case of
monomial potential $V(X)=\frac{\displaystyle X^{p+1}}{\displaystyle p+1}$,
the GKM integral is a solution of the $p$-reduced KP hierarchy. Moreover, it
satisfies, in addition, the string equation. In turn, these two conditions
fix the solution of KP hierarchy uniquely - this is exactly the GKM
partition function with monomial potential. The case of an arbitrary polynomial
potential is more complicated (and more rich). It requires the notion of the
equivalent hierarchies \cite{S} which is thoroughly discussed in Sect. 6.
We prove that the solution with polynomial potential can be generated from the
corresponding solution of the $p$-reduced KP hierarchy by the action of the
Virasoro group and is represented as another $p$-reduced $\tau$-function
corrected by the exponential factor with some quadratic form. In order to
investigate in detail the nature of the transformations between the equivalent
hierarchies one more notion is required, namely, the notion of the
quasiclassical hierarchies which is described in Sect. 7 following the
approach developed in \cite{Kri1}-\cite{TT2}. We show that the quadratic form
is related with the quasiclassical $\tau$-function. Moreover, we demonstrate
that it is possible to describe the quasiclassical hierarchy directly in
terms of GKM.
The last section contains the complete description of the GKM partition
function with an arbitrary polynomial potential.
It is proved that after redefinition of times the partition function of a
generic GKM can be presented as a product of "quasiclassical" factor and
non-deformed partition function, the latter being the solution of the
equivalent $p$-reduced KP hierarchy. We show how to extract the genuine
partition function which depends only on the sum of transformed integrable
flows and flat (quasiclassical) times and which satisfies
the standard $\L_{-p}$-constraint of the (equivalent) $p$-reduced hierarchy.

\sect{Kadomtsev-Petviashvili hierarchy}
\subsection{KP hierarchy: Lax equations}
Let $\{T\}=(T_1,T_2,\ldots,T_i,\ldots)$ be the infinite set of variables.
Consider the pseudo-differential operator (the Lax operator)
\be\label{l-op}
L\,=\,\d+\sum_{i=1}^\infty u_{i+1}(T)\d^{-i}\;;\hspace{1cm}\d=\frac{\d}{\d T_1}
\ee
where $\d^{-1}$ is a formal inverse to $\d$, i.e.
$\d^{-1}\comp\d=\d\comp\d^{-1}=1$;
for any function $f(T_1)$ and any $n\geq 1$
\be\label{k1}
\d^{-n}\comp f\,=\,\sum_{i=0}^\infty (-1)^i\frac{(n+i-1)!}{i!(n-1)!}\,
\frac{\d^if}{\d T^i_1}\comp\d^{-n-i}
\ee
Note that the Lax operator $L$ can be written as
\be\label{k2}
L\,=\,W\comp\d\comp W^{-1}
\ee
where
\be
W\,\equiv\,1+\sum_{i=1}^\infty w_i(T)\d^{-1}
\ee
and the inverse $W^{-1}$ can be calculated term by term using the Leibnitz rule
(\ref{k1}): an easy exercise gives
\be
W^{-1}\,=\,1-w_1\d^{-1}+(-w_2+w_1^2)\d^{-2}+(-w_3+2w_1w_2-w_1w_1'-w_1^3)\d^{-3}
+\ldots
\ee
where $'$ denotes the derivative w.r.t. $T_1$. Comparing (\ref{l-op}) and
(\ref{k2}) one can find the relation between functions $\{u_i\}$ and $\{w_i\}$:
\be
u_2\,=\,w_1'\\
u_3\,=\,-w_2'+w_1w_1'\\
u_4\,=\,-w_3'+w_1w_2'+w_1'w_2-w_1^2w_1'-(w_1')^2
\ee
etc.\\
Let $L^k_+$ denotes
the differential parts of the pseudo-differential operators $L^k$; for
example,
\be
L_+\,=\,\d\\
L^2_+\,=\,\d^2+2u_2\\
L^3_+\,=\,\d^3+3u_2\d+3(u_3+u_2')
\ee
etc.
We use also the notation $L^k_-$ to denote the purely pseudo-differential
part of $L^k$; evidently, $L^k=L^k_++L^k_{-}$.

\bigskip\noindent
By definition, the dependence of functions $\{u_i\}$ on
{\it time} variables $(T_1, T_2,\ldots )$ is determined by the Lax equations
\be\label{lax}
\frac{\d L}{\d T_k}\,=\,[L^k_+,L]\;;\hspace{1cm}k\geq 1
\ee
It can be shown that this set of equations is equivalent to
zero-curvature conditions
\be\label{zc}
\frac{\d L^n_+}{\d T_k}-\frac{\d L^k_+}{\d T_n}\,=\,[L^k_+,L^n_+]
\ee
The set of equations (\ref{lax}) (or, equivalently, (\ref{zc}))
is called the Kadomtsev-Petviashvili (KP) hierarchy. Let $n=2,\;m=3$ in
(\ref{zc}). Using the explicit expression for the differential polynomials
$L^2_+\,,\;L^3_+$ one can easily get the simplest equation of KP hierarchy -
the Kadomtsev-Petviashvili equation:
\be\label{k4}
\frac{\d}{\d T_1}\Big(4\frac{\d u_2}{\d T_3}-12u_2\frac{\d u_2}{\d T_1}-
\frac{\d^3 u_2}{\d T_1^3}\Big)\,-\,3\frac{\d^2 u_2}{\d T^2_2}\,=\,0
\ee

\subsection{Baker-Akhiezer functions}
Evolution equations for KP hierarchy (\ref{lax}) or (\ref{zc}) are the
compatibility conditions of the following equations:
\be\label{k5}
L\Psi\,=\,z\Psi\\
\d_{_{T_n}}\Psi\,=\,L^n_+\Psi
\ee
The function $\Psi(T,z)$ which satisfies this system is called the
Baker-Akhiezer function.
Introduce the conjugation $\d^\ast=-\d$ and put
\be\label{k6}
L^\ast\,=\,-\d+(-\d)^{-1}\comp u_2+(-\d)^{-2}\comp u_3+\ldots\\
W^\ast\,=\,1+(-\d)^{-1}\comp w_1+(-\d)^{-2}\comp w_2+\ldots
\ee
such that $L^\ast=-(W^\ast)^{-1}\comp\d\comp W^\ast$.
The adjoint Baker-Akhiezer function $\Psi^\ast(T,z)$ satisfies, by definition,
the set of equations
\be\label{k7}
L^\ast\Psi^\ast\,=\,z\Psi^\ast\\
\d_{_{T_n}}\Psi^\ast\,=\,-\,(L^n_+)^\ast\Psi^\ast
\ee
It can be shown that solutions of the systems (\ref{k6}), (\ref{k7} are
represented in the form
\be\label{k8}
\Psi(T,z)\,=\,W(T,\d)e^{\xi(T,z)}\,\equiv\,
e^{\xi(T,z)}\sum_{i=0}^\infty w_i(T)z^{-i}\\
\Psi^\ast(T,z)\,=\,W^\ast(T,\d)^{-1}e^{-\xi(T,z)}
\ee
where
\be\label{k9}
\xi(T,z)\,\equiv\,\sum_{k=1}^\infty T_kz^k
\ee
In \cite{DJKM} the following fundamental theorem has been proved:\\
Let $\Psi(T,z)$, $\Psi^\ast(T,z)$ be the Baker-Akhiezer functions of the KP
hierarchy. There exists the function $\tau(T)$ such that
\be\label{k10}
\Psi(T,z)\,=\,\frac{\displaystyle \tau(T_k-\frac{1}{kz^k})}
{\displaystyle\tau(T_k)}\;e^{\xi(t,z)}\\
\Psi^\ast(T,z)\,=\,\frac{\displaystyle \tau(T_k+\frac{1}{kz^k})}
{\displaystyle\tau(T_k)}\;e^{-\xi(t,z)}
\ee
It is not hard to see that all the functions $u_i(T)\,\;i\geq 2$ can be
represented in terms of $\tau$. For example,
\be\label{k11}
u_2\,=\, \d^2_{_{T_1}}\log\tau\\
u_3\,=\,\frac{1}{2}(\d^3_{_{T_1}}+\d_{_{T_1}}\d_{_{T_3}})\log\tau\\
u_4\,=\,\frac{1}{6}(\d^4_{_{T_1}}-3\d^2_{_{T_1}}\d_{_{T_2}}+2\d_{_{T_1}}
\d_{_{T_2}})\log\tau\,-\,(\d^2_{_{T_1}}\log\tau)^2
\ee
etc. Substitution of the first relation to (\ref{k11}) gives the
representation of the KP equation in the bilinear form
\be\label{k12}
\frac{1}{12}\tau\Big(\frac{\d^4\tau}{\d T_1^4}-4\frac{\d^2\tau}{\d T_1\d T_3}+
3\frac{\d^2\tau}{\d T^2_2}\Big)-\frac{1}{3}\frac{\d\tau}{\d T_1}
\Big(\frac{\d^3\tau}{\d T^3_1}-\frac{\d\tau}{\d T_3}\Big)+\\ +
\frac{1}{4}\Big(\frac{\d^2\tau}{\d T_1^2}+\frac{\d\tau}{\d T_2}\Big)
\Big(\frac{\d^2\tau}{\d T_1^2}-\frac{\d\tau}{\d T_2}\Big)=0
\ee
As it turns out, it is possible to rewrite all non-linear equations of
KP hierarchy as an infinite set of {\it bilinear} equations for the
$\tau$-function \cite{DJKM} in more or less compact form using the Hirota
symbols. \\
One should note that it is possible to consider more general integrable
system, namely the Toda lattice (TL) hierarchy \cite{UT} which can be thought
as a specific "gluing" of the two KP hierarchies. In this case the solutions
depend on the two infinite sets of times, $\{T_k\}$ and $\{\nT_k\}$
parametrizing the KP parts as well on the discrete time $n$ which mixes
the KP evolutions. The $\tau$-function of TL hierarchy $\tau_n(T,\nT)$ also
satisfies the infinite set of bilinear equations \cite{UT}; the simplest
evolution is described by the famous Toda equation
\be\label{toda}
\tau_n\frac{\d^2\tau_n}{\d T_1\d\nT_1}-
\frac{\d\tau_n}{\d T_1}\frac{\d\tau_n}{\d\nT_1}=\,-\,\tau_{n+1}\tau_{n-1}
\ee
The main problem is to describe the generic solutions of these
hierarchies. It will be done in section 4.

\subsection{Reduction}
The KP hierarchy is called a $p$-reduced one if for some natural $p\geq 2$
the operator $L^p$ has only differential part, i.e.
\be\label{kr1}
(L^p)_{_-}\,=\,0
\ee
In this case $L^{np}_+ =L^{np}$ for any $n\geq 1$ and from (\ref{k5}),
(\ref{k10}) it follows that
\be
\frac{\d}{\d T_{np}}\,\frac{\displaystyle \tau(T_k-\frac{1}{kz^k})}
{\displaystyle\tau(T_k)}\,=\,0
\ee
From the last relation it is clear that on the level of $\tau$-function
the condition of $p$-reduction reads
\be\label{kr2}
\frac{\d\tau(T)}{\d T_{np}}\,=\,Const\cdot\tau(T)\hspace{1cm}n=1,2,\,\ldots
\ee
Equivalently, the relations (\ref{kr2}) can be taken themselves as a
definition of $p$-reduced hierarchy.

\section{Free field realization of $gl_\infty$}
\subsection {Free fermions and vacuum states}

Let us consider the infinite set of the fermionic modes $\psi_i\,,
\;\psi^{\ast}_i\,,\;i\in\ZZ$ which satisfy the usual anticommutation
relations
\beq
\{\psi_{i},\psi^{\ast}_{j} \} =
\delta_{ij} \;\;,\hspace{0.3in} \{\psi_{i},\psi_{j} \}
=\{\psi^{\ast}_{i},\psi^{\ast}_{j} \} = 0\;\;\;\;\; i,j\in\ZZ
\eeq
Totally empty (true) vacuum $|+\infty \rangle$ is determined  by relations
\be
\psi_{i}\;|+\infty \rangle = 0 \;\; , \;\; i \in \ZZ
\label{eq:a}
\ee
Then the n-th "vacuum" state $|n\>$ is defined as follows:
\be
|n \> = \psi^{\ast}_{n} \psi^{\ast}_{n+1} \ldots |+\infty\>
\ee
thus satisfying the conditions (which themselves can be taken
as definition of such state):
\be\label{vac1}
\psi^{\ast}_{k}|n \> = 0 \;,\;\;\;\;k\geq n \;\;;\;\;\;
\;\;\;\;\;\; \psi_{k}|n \> = 0 \;\;\;\;\; k<n
\ee
Similarly, the left (dual) $n$-th vacuum $\<n|$ is defined by
conditions
\be\label{vac2}
\<n|\psi^{\ast}_{k} = 0 \;,\;\;\;\;k < n \;\;;\;\;\;
\;\;\;\;\;\; \<n|\psi_{k} = 0 \;,\;\;\;\; k\geq n
\ee
One can select the particular state, for example, $|0\>$ and
consider the normal ordering of the fermions with respect to this
preferred vacuum.
In this case the annihilation operators are $\psi_i\;,\;\; i<0$ and
$\psi^{\ast}_i\;,\;\; i\geq0$ and, therefore, the normal ordering
is defined as follows:
\be\label{ord1}
\psi_i\psi^{\ast}_j = \;:\psi_i\psi^{\ast}_j: + \theta(-i-1)\delta_{ij}
\ee

\subsection{Boson-fermion correspondence}
It is convenient to introduce the free fermionic fields
\beq
\psi(z) \equiv \sum_{i \in \z} \psi_{i}z^{i}\;\;\;,\;\;\;\psi^{\ast}(z)
\equiv \sum_{i \in \z} \psi^{\ast}_{i}z^{-i}
\eeq
which, in turn, can be expressed in terms of the free {\it bosonic}
field $\varphi(z)$
\be\label{eq:phi}
\varphi(z) = q - i p \log z +  i \sum_{k \in \z} \frac{J_{k}}{k}z^{-k} \\

[q,p]= i\;\; ; \;\;\;\;\; [J_{m},J_{n}]=m\delta_{m+n,0}
\ee
according to well known formulae
\be \label{rep1}
\psi(z)= \; :e^{{
i \varphi(z)}}: \;\equiv \\
\equiv e^{i q}\;e^{ p\log z}\;
\exp \left(\sum_{k=1}^{\infty}\frac{J_{-k}}{k}z^{k}\right) \times \exp
\left(-\sum_{k=1}^{\infty}\frac{J_{k}}{k}z^{-k}\right)\;\; ,
\label{eq:q}
\ee
\bigskip
\be \label{rep2}
\psi^{\ast}(z)= z\;:e^{{
- i \varphi(z)}}:\; \equiv \\
\equiv ze^{-i q}\;e^{-p\log z}\;
\exp \left(-\sum_{k=1}^{\infty}\frac{J_{-k}}{k}z^{k}\right)
\times \exp\left(\sum_{k=1}^{\infty}\frac{J_{k}}{k}z^{-k}\right)
\ee
Note, that under the formal Hermitian conjugation
\be\label{con}
(z)^\dagger = z^{-1}\;\,; \ \ \ \ (J_k)^\dagger = J_{-k}\\
(q)^\dagger = q\;\,; \ \ \ \ \;(p)^\dagger = p
\ee
we have the involution
\be
(\psi(z))^\dagger = \psi^{\ast}(z)\;\; ,
\;\;\;\;\;\;(\psi^{\ast}(z))^\dagger = \psi(z)
\ee
It can be shown that vacua $|n\>$ are eigenfunctions of the operator
$p$:
\be \label{p}
p|n\> = n|n\>\;;\hspace{0.5cm}\<n|p=n\,\<n|
\ee
and zero bosonic mode shifts the vacua, i.e. changes its  charge
\be\label{q}
\left\{
\begin{array}{l} 
e^{imq}|n\> = |n+m\>\;\;\\
\<n|e^{imq} = \<n-m|\;\;
\end{array}\right.
  \;\;\;\;\;\; m\in\ZZ\;\; .
\ee
Using the definition (\ref{eq:phi}) one can show that
\be\label{ord2}
:e^{{
i \alpha \varphi(z)}}:\;
:e^{{
i \beta\varphi(w)}}:\; =
(z-w)^{\alpha \beta}:e^{{
i \alpha \phi(z)+ i \beta \phi(w)}}:
\ee
and, therefore,
\be\label{ord}
\psi(z) \psi^{\ast}(w) = \frac{w}{z-w}:e^{{
i \varphi(z)-i\varphi(w)}}:\; \equiv \\
\equiv \;:\psi(z) \psi^{\ast}(w): + \frac{w}{z-w}
\ee
The last expression being expanded near the point $w \sim z$
enables to rewrite the bosonic field via the fermionic ones:
\be
i\d_z\varphi(z) = \frac{1}{z}\;:\psi(z) \psi^{\ast}(z):\;=\;
\sum_{k\in\z}J_kz^{-k-1}
\ee
or, equivalently, the bosonic currents can be represented as bilinear
combination of the fermionic modes:
\be\label{curr}
J_k = \sum_{i\in\z}:\psi_i \psi^{\ast}_{i+k}: \;, \;\;\;\; k\in\ZZ
\ee
Obviously, the normal ordering in (\ref{curr}) is essential
only for $J_0\equiv p$. Using (\ref{curr}) it is easy to see that
\be \label{curr2}\left\{
\begin{array}{l}
J_k|n\> \equiv 0\\
\<n|J_{-k} \equiv 0
\end{array}\right.
\;\;\;\;\;\;k > 0\;\; ,\;\;\;\;n\in\ZZ
\ee
One should mention that not only the bosonic currents can be expressed as
bilinear combination of the free fermions. Actually, this is true for the
whole family of $gl_\infty$ generators (sometimes called the
$W_{1+\infty}$-generators); for example, one can derive
analogous boson-fermion corresponding for the Virasoro generators:
\be\label{Vir}
\L_k\,\equiv\,\frac{1}{2}\sum_{i\in\z}:J_iJ_{k-i}:\,=\,
\sum_{i\in\z}\Bigl(i +\frac{k+1}{2}\Bigr):\psi_i\psi^{\ast}_{i+k}:
\ee
The bosonization formulae are very useful tool to calculate
different correlators containing the fermionic operators.

\section{$\tau$-functions in free field representation}
In this section the solutions of KP (more generally, Toda) hierarchy are
represented in the form of the fermionic correlators parametrized by the
infinite set of continuous variables. The fermionic language is very
convenient for the integrable systems since it enables to represent an
arbitrary solution in the specific determinant form. This, in turn, allows to
identify the GKM partition function with appropriate solution of the
hierarchy.

\subsection{Fermionic correlators, Wick theorem and solution of KP (TL)
hierarchy}
Let us introduce the "Hamiltonians"
\be\label{hm}
H(T) \equiv \sum_{k=1}^{\infty} T_{k}J_{k}\,,\hspace{1cm}
\nH(\nT) \equiv \sum_{k=1}^{\infty} \nT_{k}J_{-k}
\ee
where $\{T_k\}$ and $\{T_k\}$ are the infinite sets of parameters
(sometimes called the sets of positive and negative times respectively).
We define the fermionic correlators ($\tau$-functions) with the following
parameterization by these times
\be\label{tauTL}
\tau_n(T,\nT|g) = \< n|e^{H(T)}ge^{-\nH(\nT)}|n \>\,\equiv\,\< n|g(T,\nT)|n \>
\ee
where
\be\label{point}
g=\; :\exp \Bigl\{\sum_{i,j \in \z} A_{ij}\psi_{i} \psi^{\ast}_{j}\Bigr\}:
\ee
with $||A_{ij}|| \in gl_\infty$. In the most cases we shall write
$\tau_n(T,\nT)$ for brevity.
We assume that the (infinite)
matrix $||A_{ij}||$ satisfies such a requirements that the correlator
(\ref{tauTL}) is well defined. As an example, the matrix with almost all
zero entries is suitable. The wide class of the suitable matrices are the
Jacobian ones: $A_{ij}=0$ for $|i-j|\gg 1$. The more general conditions can
be found in \cite{SP}. One should mention that
the normal ordering in (\ref{point}) is taken with respect the zero vacuum
state $|0\>$ (see (\ref{ord1})); it is equivalent to (\ref{ord}).\\
Note also
that every element of the type (\ref{point}) rotates the fermionic modes:
\be\label{ro1}
g\psi_ig^{-1}\,=\,R_{ki}\psi_k\;;\hspace{0.6cm}
g\psi^\ast_ig^{-1}\,=\,R^{-1}_{ik}\psi^\ast_k
\ee
with some (infinite) matrix $||R||\in GL_\infty$. As an example, the
exponentials containing the Hamiltonians give the transformations
\be\label{ro2}
\begin{array}{ll}
e^{H(T)}\psi(z)e^{-H(T)}\,=\,e^{\xi(T,z)}\psi(z)\;\,;&\hspace{0.7cm}
e^{H(T)}\psi^\ast(z)e^{-H(T)}\,=\,e^{-\xi(T,z)}\psi^\ast(z)\\
e^{\nH(\nT)}\psi(z)e^{-\nH(\nT)}\,=\,e^{\xi(\nT,z^{-1})}\psi(z)\;\,;
                &\hspace{0.7cm}
e^{\nH(\nT)}\psi^\ast(z)e^{-\nH(\nT)}\,=\,e^{-\xi(\nT,z^{-1})}\psi^\ast(z)
\end{array}
\ee
because of commutator relations $[J_k,\psi(z)]=z^k\psi(z)\,,\;\,
[J_k,\psi^\ast(z)]=z^{-k}\psi^\ast(z)$ (the latter are simple consequence
of the fermionic representation (\ref{curr2})).

\bigskip\noindent
The fermionic correlators introduced above have a very specific dependence
on the infinite sets of times $\{T_k\},\,\{\nT_k\}$.
The main statement is that the correlators (\ref{tauTL}) solve the
Toda lattice hierarchy; in particular, as a function of the positive times
$\{T_k\}$ these correlators are solutions of the KP hierarchy: each
particular solution is parametrized by the given matrix $||A_{ij}||$. It can
be proved in full generality using the so-called bilinear identity
\cite{DJKM}. For the local purposes it is enough, however, to show that the
simplest equations of the mentioned hierarchies are satisfied.
It is possible to deduce them starting directly from the fermionic correlators.
Note that we shall deal with the only KP hierarchy in what follows.
Nevertheless, as an instructive example, let us derive 2-dimensional Toda
equation which is the first equation of the Toda hierarchy. The example
shows the natural appearance of the determinant representations in the
context of the integrable systems; besides, the similar technique will be used
below quite extensively.

\bigskip\noindent
All the correlators similar to (\ref{tauTL})
are expressed in terms of the free fields, hence, the Wick theorem is
applicable; as an example
\be
\frac{\<n|\psi_{i_1}\ldots\psi_{i_k}g(T,\ov{T})
\psi^{\ast}_{j_1}\ldots\psi^{\ast}_{j_k}|n\>}{\<n|g(T,\ov{T})|n\>}\;=\;
\left.\det\frac{\<n|\psi_{i_a}g(T,\ov{T})\psi^{\ast}_{j_b}|n\>}
{\<n|g(T,\ov{T})|n\>}\right|_{a,b=1}^k
\ee
This key observation gives easy way to prove that the $\tau$-function
(\ref{tauTL}) satisfies the standard Toda equation. Indeed, using
the fermionic representation (\ref{curr}) of the currents $J_k$
together with the
definition of the vacuum states (\ref{vac1}), (\ref{vac2}) one gets
\be
\d_{_{T_1}}\d_{_{\nT_1}}\tau_n = \,-\;\<n|J_1e^{H(T)}ge^{-\nH(\nT)}J_{-1}|n \>
=\\ =\,-\;\<n|\psi_{n-1}\psi^{\ast}_ng(T,\ov{T})\psi_n\psi^{\ast}_{n-1}|n \>

\ee
Using the Wick theorem, this expression can be written in the form
\be\label{eqTL}
\d_{_{T_1}}\d_{_{\nT_1}}\tau_n  = \,-\;\frac{1}{\tau_n}
\Bigl\{\<n|\psi_{n-1}\psi^{\ast}_ng(T,\ov{T})|n \>
\<n|_ng(T,\ov{T})\psi_n\psi^{\ast}_{n-1}|n \> +\\
+ \<n|\psi_{n-1}g(T,\ov{T})\psi^{\ast}_{n-1}|n \>
\<n|\psi^{\ast}_ng(T,\ov{T})\psi_n|n \>\Bigr\}\;\; .
\ee
Recalling the definitions again one can rewrite every term in the last
formula in terms of the $\tau$-functions and their derivatives; namely,
\be
\begin{array}{ll}
\<n|\psi_{n-1}g(T,\ov{T})\psi^{\ast}_{n-1}|n \> = \tau_{n-1}\;\; ,\;\;\;\;\;\;
& \<n|\psi^{\ast}_ng(T,\ov{T})\psi_n|n \> = \tau_{n+1}\;\; ,\\
\<n|\psi_{n-1}\psi^{\ast}_ng(T,\ov{T})|n \> = \d_{_{T_1}}\tau_n\;\; ,
\;\;\;\;\;\; &\<n|g(T,\ov{T})\psi_n\psi^{\ast}_{n-1}|n \> =
\,-\;\d_{_{\nT_1}}\tau_n
\end{array}
\ee
and, therefore, (\ref{eqTL}) reduces to Toda equation
\be\label{TTT}
\d_{_{T_1}}\d_{_{\ov{T}_1}}\log\,=\,-\,\frac{\tau_{n+1}\tau_{n-1}}{\tau^2_n}
\ee
which is equivalent to (\ref{toda}).
The analogous (though more involved) calculations show that $\tau_n$
as a function of the positive times $T_1,T_2,T_3$ satisfies the
Kadomtsev-Petviashvili equation (\ref{k12}) for any fixed $n$. Let us stress
again that the complete list of bilinear equations for the $\tau$-functions
is represented in \cite{DJKM, UT}.

\subsection{Determinant representation of $\tau$-functions}
Here we represent an arbitrary solution of the KP hierarchy in the determinant
form which is crucial in what follows.
Let us calculate the fermionic correlator
$\<n+N|\psi(\mu_N)\ldots\psi(\mu_1)g|n\>$ in two different ways.
First of all, using the definition of the vacua and applying the Wick theorem,
the correlator can be written in the determinant form:
\be\label{start1}
\<n+N|\psi(\mu_N)\ldots\psi(\mu_1)g|n\>\,=\,
\<n|\psi_n^\ast\ldots\psi^\ast_{n+N-1}\psi(\mu_N)\ldots\psi(\mu_1)g|n\>\,=\\
=\,\<n|g|n\>\,\det\frac{\<n|\psi^\ast_{n+i-1}\psi(\mu_j)g|n\>}{\<n|g|n\>}
\ee
On the other hand, using the boson-fermion correspondence (\ref{eq:q}),
the normal ordering (\ref{ord2}), and the formulas (\ref{p}), (\ref{q}),
(\ref{curr2}) describing the action of the different operators on the
vacuum state $\<N|$, one can write
\be\label{start2}
\<n+N|\psi(\mu_N)\ldots\psi(\mu_1)g|n\>\equiv
\Delta(\mu)\<n+N|:\exp\Big\{i\sum_{j=1}^N\phi(\mu_j)\Big\}:g|n\>\;=\\
=\,\Delta(\mu)\prod_{j=1}^N\mu^n_j\,\<n|\exp\Big\{\sum_{k=1}^\infty
T_kJ_k\Big\}g|n\>
\ee
where in the r.h.s. the $\tau$-function appears with the specific
parametrization of the positive times
\be\label{mi}
T_k\,\equiv\,-\,\frac{1}{k}\sum_{j=1}^N\mu_j^{-k}
\ee
The parametrization (\ref{mi}) has been introduced in \cite{Mi}.
We shall call such
representation of times the Miwa parametrization (respectively, the set
$\{\mu_i\}$ is called the Miwa variables). Note that for $N$ finite only first
$N$ times $T_1,\ldots ,T_N$ are functionally independent. Equivalently, only
first $N$ equations of the KP hierarchy have a non-trivial sense
(all higher equations are functionally dependent on the first $N$
ones). We shall deal with such restricted hierarchy in what follows.
Comparing the relations (\ref{start1}), (\ref{start2}) one arrives
to the following statement. For any finite $N$
the $\tau$-functions of the KP hierarchy being written in the Miwa
variables (\ref{mi}) can be represented in the determinant form
\be\label{det}
\tau_n(T)\,=\,\<n|g|n\>\,\frac{\det\,\phi^{(can)}_i(\mu_j)|_{i,j=1}^N}
{\Delta(\mu)}
\ee
where {\it the canonical basis vectors}
\be\label{bv}
\phi^{(can)}_i(\mu)\,=\,\mu^{-n}\,
\frac{\<n|\psi^\ast_{n+i-1}\psi(\mu)g|n\>}{\<n|g|n\>}
\hspace{1cm}i=1,\,2,\,\ldots
\ee
have the following asymptotics
\be\label{as1}
\phi^{(can)}_i(\mu)\,=\,\mu^{i-1}+O\Big(\frac{1}{\mu}\Big)
\hspace{1cm}\mu\to\infty
\ee
Moreover, the opposite statement is true. Namely, any functions
$\tau(\mu_1,\ldots,\mu_N)$ of the form
\be\label{det2}
\tau(T)\,=\,\frac{\det\,\phi_i(\mu_j)}{\Delta(\mu)}\;;
\hspace{1.5cm}T_k\,\equiv\,-\,\frac{1}{k}\sum_{j=1}^N\mu_j^{-k}
\ee
whose basis vectors $\phi_i(\mu)\,,\;i=1,\,2,\,\ldots$ have
the asymptotics
\be\label{as2}
\phi_i(\mu)\,=\,\mu^{i-1}\Big(1+O\Big(\frac{1}{\mu}\Big)\Big)
\hspace{1cm}\mu\to\infty
\ee
solve the KP hierarchy.
The set $\{\phi_i(\mu)\}$ satisfying the asymptotics (\ref{as2}) is naturally
identified with the projective coordinates of a point of
Grassmannian \cite{SP}.
More precisely, the vectors $\{\phi_i(\mu)\}$
can be transformed to the canonical ones taking the appropriate linear
combinations (clearly, such transformation does not change the determinant in
(\ref{det2})). Then, there exists the element (\ref{point})
of the Grassmannian such that
transformed basis vectors can be written as a fermionic correlators
(\ref{bv}) (for some fixed $n$) and, consequently, $\tau(\mu_1,\ldots,\mu_N)$
have the form (\ref{tauTL}) in the Miwa parametrization (\ref{mi}).
To summarize, any infinite set of the vectors (\ref{as2}) describes
the particular solution of KP hierarchy via determinant form (\ref{det2}).

\subsection{Time derivatives}\label{td}
Let us find the expression of the time derivatives $\d\tau/\d{T_k}$
for the $\tau$-function written in the determinant form (\ref{det}).
As in (\ref{mi}), we assume the finite number $N$ of the
Miwa variables. Hence, only first $N$ times $T_k$ are functionally
independent and all formulas below have a sense for $\d\tau/\d{T_1},\ldots ,
\d\tau/\d{T_N}$ only. From (\ref{start2})
\be\label{1}
\frac{\d\tau_n}{\d T_k}\,=\,\frac{\prod\mu_i^{-n}}{\Delta(\mu)}\,
\<n+N|\psi(\mu_N)\ldots\psi(\mu_1)J_kg|n\>\,\equiv\\
\equiv\frac{\prod\mu_i^{-n}}{\Delta(\mu)}\Big\{
\<n+N|J_k\psi(\mu_N)\ldots\psi(\mu_1)g|n\>+
\sum_{i=1}^N\<n+N|\psi(\mu_N)\ldots[\psi(\mu_i),J_k]\ldots\psi(\mu_1)g|n\>
\Big\}
\ee
Since the currents $J_k=\sum_{j\in\z}\psi_j\psi_{j+k}^\ast$ satisfy the
commutation relations $[J_k,\psi(\mu)]=\mu^k\psi(\mu)$ the last expression
can be written in the form
\be\label{dk1}
\frac{\d\tau_n}{\d T_k}\,=\,
\frac{\prod\mu_i^{-n}}{\Delta(\mu)}\,
\<n|\psi_n^\ast\ldots\psi_{n+N-1}^\ast
\Big\{\sum_{j=n+N-k}^{n+N-1}\psi_j\psi_{j+k}^\ast\Big\}\psi(\mu_N)\ldots
\psi(\mu_1)g|n\>-\tau_n(x)\sum_{i=1}^N\mu_i^k
\ee
where, according to the definition of the vacua (\ref{vac2}), the
action of $J_k$ on the state $\<n+N|$ reduces to the action of the
finite number of the fermionic modes with $n+N-k\le j\le n+N-1$. This fact
allows to represent the expression (\ref{dk1}) in the compact determinant
form. Indeed, since $j\ge n+N-k$ and $k\le N$ (i.e. $j\ge n$) it is clear
that $\<n|\psi_j\psi^\ast_{j+k}=0$ and the moving of the operator
$\sum_{j=n+N-k}^{n+N-1}\psi_j\psi_{j+k}^\ast$ to the left state results
to appropriate shifts of the modes $\psi^\ast_n,\ldots , \psi^\ast_{n+N-1}$.
For example, for $k=1$ one gets the only correlator
$\<n|\psi^\ast_n\ldots\psi^\ast_{n+N-2}\psi^\ast_{n+N}g\psi(\mu_n)
\ldots\psi(\mu_1)g|n\>$ and, therefore, the first term in (\ref{dk1}) has the
determinant form similar to (\ref{det}) (with the shifted last row
$\phi^{(can)}_N\,\to\,\phi^{(can)}_{N+1}$). It is evident that for arbitrary
$k\leq N$ the first term in (\ref{dk1}) can be represented
as the sum of the shifted determinants
\be\label{dder}
\frac{\<n|g|n\>}{\Delta(\mu)}\,\sum_{m=1}^N\,\left|
\begin{array}{ccc}
\phi^{(can)}_1(\mu_1) &\ldots & \phi^{(can)}_1(\mu_N)\\
\ldots &\ldots&\ldots \\
\phi^{(can)}_{m-1}(\mu_1) &\ldots&\phi^{(can)}_{m-1}(\mu_N)\\
\phi^{(can)}_{m+k}(\mu_1) &\ldots&\phi^{(can)}_{m+k}(\mu_N)\\
\phi^{(can)}_{m+1}(\mu_1) &\ldots&\phi^{(can)}_{m+1}(\mu_N)\\
\ldots &\ldots&\ldots \\
\phi^{(can)}_{N}(\mu_1) &\ldots&\phi^{(can)}_{N}(\mu_N)
\end{array}\right|
\ee
Hence, one arrives to the following formula:
\be\label{3der}\hspace{-0.5cm}
\frac{\d}{\d T_k}\!\left(\frac{\det \phi^{(can)}_i(\mu_j)}{\Delta(\mu)}\right)
\!=\!\frac{1}{\Delta(\mu)}\!\sum_{m=1}^N\left|
\begin{array}{ccc}
\phi^{(can)}_1(\mu_1) &\ldots & \phi^{(can)}_1(\mu_N)\\
\ldots &\ldots&\ldots \\
\phi^{(can)}_{m-1}(\mu_1) &\ldots&\phi^{(can)}_{m-1}(\mu_N)\\
\!\!\phi^{(can)}_{m+k}(\mu_1)\!-\!\mu^k_1\phi^{(can)}_{m}(\mu_1) &\ldots &
\!\!\phi^{(can)}_{m+k}(\mu_N)\!-\!\mu^k_N\phi^{(can)}_{m}(\mu_N)\\
\phi^{(can)}_{m+1}(\mu_1) &\ldots&\phi^{(can)}_{m+1}(\mu_N)\\
\ldots &\ldots&\ldots \\
\phi^{(can)}_{N}(\mu_1) &\ldots&\phi^{(can)}_{N}(\mu_N)
\end{array}\right|
\ee
Introducing the formal operator which shifts the indices
of the canonical basis vectors
\be\label{sh}
B(\mu)\phi^{(can)}_i(\mu)\,\equiv\, \phi^{(can)}_{i+1}(\mu)
\ee
one can write the final answer in more compact notations:
\be\label{time}
\frac{\d}{\d T_k}\,\left(\frac{\det \phi^{(can)}_i(\mu_j)}{\Delta(\mu)}\right)
\,=\,\frac{1}{\Delta(\mu)}
\sum_{m=1}^N\Big(B^k(\mu_m)-\mu_m^k\Big)\,\det\,\phi^{(can)}_i(\mu_j)
\ee
This is the first important formula we need in what follows. As an
immediate application one can consider the translation of the notion
of $p$-reduced KP hierarchy to the language of the Grassmannian.
Suppose that for some natural $p>1$ the quantity $\mu^p\phi^{(can)}_m(\mu)$
can be expanded in the canonical basis vectors, i.e. for any $m\geq 1$
\be\label{p-r}
\mu^p\phi^{(can)}_m(\mu)\subset\mbox{Span}\,\{\phi^{(can)}(\mu)\}
\ee
Writing
\be
\phi^{(can)}_m(\mu)\,\equiv\,\mu^{m-1}+\sum_{j=1}^\infty\al_{mj}\mu^{-j}
\ee
it is easy to see that
for any $n\geq 1$ the following expansion holds:
\be
\mu^{np}\phi^{(can)}_m(\mu)\,=\,\phi^{(can)}_{m+np}(\mu)+
\sum_{j=1}^{np}\al_{mj}\phi^{(can)}_{np-j+1}
\ee
Due to determinant structure in (\ref{3der}) every row containing the terms
$\phi^{(can)}_{m+np}-\mu^{np}\phi^{(can)}_m$ gives non-trivial
contribution $-\al_{m,np-m+1}\phi^{(can)}_m\,;\;\;1\leq m\leq np$
(provided $np\leq N$),
hence,
\be\label{p-r2}
\frac{\d\tau(T)}{\d
T_{np}}\,=\,-\,\tau(T)\sum_{m=1}^{np}\al_{m,np-m+1}\;;\hspace{1cm}np\leq N
\ee
assuming that (\ref{p-r}) holds. In the limit $N\to\infty$ this is exactly
the case of $p$-reduced KP hierarchy. Hence, the conditions (\ref{p-r}) and
(\ref{p-r2}) are equivalent \cite{SP}.

\subsection{Action of the Virasoro generators}\label{virs}
Literally the same calculation can be performed for any $W$-generators.
Consider, for example, the Virasoro generators
\be\label{TVir}
\L_k(T)\,=\,\frac{1}{2}\sum_{a+b=-k}\!\!abT_aT_b+
\sum_{a-b=-k}\!\!\!aT_a\frac{\d}{\d T_b}+
\frac{1}{2}\sum_{a+b=k}\frac{\d^2}{\d T_a\d T_b}
\ee
then, evidently,
\be
\L_k(T)\tau_n(T)\,=\,\<n|e^{H(T)}\,\L_k(J)\,g|n\>
\ee
where the fermionic Virasoro generators $\L_k(J)$ (\ref{Vir}) satisfy the
commutations relations
\be\label{vir2}
[\,\L_k(J),\psi(\mu)]\,=\,\Big(\mu^{k+1}\frac{\d}{\d\mu}+
\frac{k+1}{2}\mu^k\Big)\psi(\mu)\,\equiv\,A_k(\mu)\psi(\mu)
\ee
Consider the subset $\{\L_{-k}(J)\,,\;k>0\}$.
Taking into account the identity $\<n+N|\L_{-k}(J)=0\,,\;k>0$ one gets
instead of (\ref{1})
\be\label{Vir3}
\L_{-k}(T)\tau_n(T)\,=\,-\,
\frac{\prod\mu_i^{-n}}{\Delta(\mu)}\,
\sum_{m=1}^N\<n+N|\psi(\mu_N)\ldots[\psi(\mu_m),L_{-k}(J)]
\ldots\psi(\mu_1)g|n\>\,=\\=\,-\,
\frac{\prod\mu_i^{-n}}{\Delta(\mu)}\,
\sum_{m=1}^NA_{-k}(\mu_m)\,\<n+N|\psi(\mu_N)\ldots\psi(\mu_1)g|n\>\,=\\
=\,nkT_k\tau_n(T)\,-\,\frac{\<n|g|n\>}{\Delta(\mu)}\,\sum_{m=1}^N\,
A_{-k}(\mu_m)\,\det\,\phi^{(can)}_i(\mu_j)
\ee
In particular, the standard $\tau$-function of the KP hierarchy
$\tau_{n=0}(T)\equiv\tau(T)$ satisfies the relation
\be\label{Vir4}
\L_{-k}(T)\,\left(\frac{\det \phi^{(can)}_i(\mu_j)}{\Delta(\mu)}\right)
\,=\,-\,\frac{1}{\Delta(\mu)}\,\sum_{m=1}^N\,
A_{-k}(\mu_m)\,\det\,\phi^{(can)}_i(\mu_j)\\
A_{-k}(\mu)\,=\,\mu^{1-k}\frac{\d}{\d\mu}+
\frac{1-k}{2}\mu^{-k}
\ee
(we shall see below that the GKM partition function
corresponds exactly to the choice of $0$-vacuum state).

\bigskip\noindent
Similarly to (\ref{p-r}) consider the case when for some $q>1$
\be\label{vr1}
A_{-q}(\mu)\phi^{(can)}_i(\mu)\subset\mbox{Span}\,\{\phi^{(can)}(\mu)\}
\ee
From (\ref{Vir4}) it follows that the solution of the KP hierarchy is
invariant w.r.t. action of the corresponding Virasoro generator:
\be\label{vr2}
\L_{-q}(T)\tau(T)\,=\,0
\ee
In the next subsection it will shown that the GKM partition function
satisfies the conditions quite similar to (\ref{p-r2}) and (\ref{vr2}).

\bigskip\noindent
Relations (\ref{time}) and (\ref{Vir4}) are the simplest examples of
$W$-generators acting on $\tau$-functions in the Miwa parametrization.
Using the fermionic representation it is possible to write down
the similar expressions for the higher generators.

\section{Generalized Kontsevich model: Preliminary investigation}
\subsection{GKM: the definition}
Recall that the standard Hermitian one-matrix model is defined as a multiple
integral over $n\times n$ Hermitian matrix $X$
\be\label{1m}
Z_n[t]\,=\,\int e^{-\Tr\,S(X,t)}dX
\ee
where the action $S(X,t)$ depends on infinitely many coupling constants
("the times")
\be
S(X,t)\,=\,\sum_{k=1}^\infty t_kX^k
\ee
and the measure
\be\label{me}
dX\,=\,\prod_{i=1}^{n}dX_{ii}\,
\prod_{i<j}2\,d(\mbox{Re}X_{ij})d(\mbox{Im}X_{ij})
\ee
is chosen in such a way that the following normalization condition is
fixed:
\be
\int e^{-\frac{1}{2}\Tr\,X^2}dX\,=\,(2\pi)^{n^2/2}
\ee
After the integration over the angle variables \cite{Mehta}
the partition function (\ref{1m}) results to $n$-tuple integral over the
eigenvalues $x_1,\ldots ,x_n$ of the matrix $X$:
\be\label{2m}
Z_n[t]\,=\,\frac{(2\pi)^{\frac{n(n-1)}{2}}}{\prod_{k=1}^n k!}\,\int
\Delta^2(x)\,\prod_{i=1}^n e^{-S(x_i,t)}dx_i
\ee
where
\be\label{vdm}
\Delta(x)\,\equiv\,\prod_{i>j}(x_i-x_j)
\ee
is the van der Monde determinant and
\be
U_n\,\equiv\,\frac{(2\pi)^{\frac{n(n-1)}{2}}}{\prod_{k=1}^n k!}
\ee
is a volume of the group $SU(n)$. The partition function (\ref{2m}) possesses
a remarkable integrability property: as a function of times $\{t_k\}$ and
discrete variable $n$ (the size of the matrix) it is a
solution of so-called Toda chain hierarchy. In particular, the function
\be
\tau_n(t)\,\equiv\,\frac{1}{n!U_n}\,Z_n[t]
\ee
satisfies the famous Toda equation
\be
\frac{\d^2\log\tau_n}{\d t_1^2}\,=\,\frac{\tau_{n+1}\tau_{n-1}}{\tau^2_n}
\ee

\bigskip\noindent
The main object we shall discuss below is quite different one-matrix integral
depending on the external $N\times N$ Hermitian matrix $M$:
\be\label{def}
Z^V_N[M]\,=\,\frac{\int e^{-S(M,Y)}dY}{\int e^{-S_2(M,Y)}dY}
\ee
where the measure is the same as in (\ref{me}) (with $n$ substituted by $N$).
The explicit dependence on the matrix $M$ comes from the action $S(M,Y)$
and its quadratic part $S_2(M,Y)$; for any Taylor series
$V(Y)$ we set, by definition,
\be
S(M,Y)\,=\,\mbox{Tr}\,\big[V(Y+M)-V'(M)Y-V(M)\big]
\ee
such that this action does not contain the constant and linear terms in $Y$.
The denominator in (\ref{def}) is interpreted as a natural normalization factor
and is nothing but a Gaussian integral determined by the
quadratic part of the original action:
\be\label{qv}
S_2(M,Y)\,=\,\lim_{\e\to 0}\,\frac{1}{\e^2}\,S(M,\e Y)
\ee

\bigskip\noindent
It is clear that the integral (\ref{def}) depends only on the eigenvalues
$\mu_{_1},\,\ldots\,,\mu_{_N}$ of the external matrix $M$. It is more
reasonable, however, to use another parametrization of the partition function
$Z^V_N$ treating it as a function of {\it the time variables} $T_k$ defined by
relations
\be\label{3m}
T_k\,=\,-\,\frac{1}{k}\,\sum_{i=1}^N\mu_i^{-k}
\ee
- these are appropriate analogues of times entering in a definition of the
standard matrix model (\ref{1m})
\footnote{Nevertheless, to write the explicit expression of the partition
function in times $\{T_k\}$ requires some additional job.}.
The appearance of such variables is very natural by the reasons discussed
below.

\bigskip\noindent
The matrix model (\ref{def}) is called the Generalized Kontsevich Model.
The reason for this is that for the special choice of potential
\be
V(Y) = Y^3/3
\ee
the integral (\ref{def}) becomes the partition function of original Kontsevich
model \cite{Kon}:
\be\label{4m}
Z^{(2)}_N[M]\,=\,{\int dY\ e^{-1/3\ TrY^3 - TrMY^2}\over
\int dY\ e^{-TrMY^2}}
\ee
Expression (\ref{4m}) has been derived in \cite{Kon} as a representation of
the generating
functional of intersection numbers of the stable cohomology classes on the
universal moduli space, $i.e$. it is defined to be a partition function of
Witten's $2d$ topological gravity \cite{Wit90}. In \cite{MMM91b}, (see also
\cite{MS91,Wit91} for
alternative derivations) it was shown that as $N \rightarrow  \infty \ $
$Z^{(2)}_\infty $ considered as a function of time variables (\ref{4m}),
satisfies
the set of Virasoro constraints
\be\label{5m}
\L^{(2)}_nZ^{(2)}_\infty  = 0, \ \ \ n \geq -1\
\ee
\be
\L^{(2)}_n= {1\over 2} \sum _{k\ odd} kT_k\partial /\partial T_{k+2n}+
{1\over 4}
\sum _{{a+b=2n}\atop {a,b\ odd\ and>0}}\partial ^2/
\partial T_a\partial T_b+\nn \\
+ {1\over 4}
\sum _{{a+b=-2n}\atop {a,b\ odd\ and>0}}aT_abT_b+ {1\over 16}\delta _{n,0} -
\partial /\partial T_{3+2n}.
\ee
Constraints (\ref{5m}) are exactly the equations
\cite{FKN1},\cite{DVV}, imposed on  the square root of
the partition function (\ref{1m}) in the double-scaling limit

\subsection{GKM in the determinant form}
After the shift of the integration variable
\be
X\,=\,Y+M
\ee
the numerator in (\ref{def}) can be written in the form
\be\label{d1}
\int e^{-S(Y,M)}dY\,=\,e^{\Tr\,[V(M)-MV'(M)]}\,F [V'(M)]
\ee
where
\be\label{d2}
F[\Lambda]\,=\,\int e^{-\Tr\,V(X)+\Tr\,\Lambda X}dX\;;
\hspace{1cm}\Lambda\,\equiv\,V'(M)
\ee
Using the integration over the angular variables of the matrix $X$ according
to \cite{IZ},\cite{CMM}, one gets
\be\label{d3}
F[\Lambda]\,=\,(2\pi)^{N(N-1)/2}\frac{1}{\Delta(\la)}
\int\,\Delta(x)\prod_{i=1}^N e^{-V(x_i)+\la_ix_i}dx_i
\ee
where $\{\la_i\}$ and $\{x_i\}$ are eigenvalues of the matrices $\Lambda$
and $X$ respectively. Therefore, the function $F[V'(M)]$ in (\ref{d2}) can
be represented as
\be\label{d4}
F[V'(M)]\,\sim\,
\left.\frac{1}{\Delta(V'(\mu))}\,\det\,\Big\{\int\,x^{j-1}
e^{-V(x)+V'(\mu_i)x}dx\Big\}\right|_{i,j=1}^N
\ee
where $\Delta(V'(\mu))\equiv\prod_{i>j}(V'(\mu_i)-V'(\mu_j))$ in accordance
with the definition (\ref{vdm}) and the unessential constant factor is
omitted.

\bigskip\noindent
Proceed now to the denominator of (\ref{def})
\be\label{d5}
D^V_N[M]\,\equiv\,\int dY\ e^{-S_2(M,Y)}
\ee
Making use of $SU(N)$-invariance of the measure $dY$ one can easily diagonalize
$M$ in (\ref{d5}). Of course, this does not imply any integration over angular
variables and provide no factors like  $\Delta (Y)$. Then for evaluation of
(\ref{d5}) it remains to use the obvious rule of Gaussian integration,
\be
\int dY\ e^{-\sum ^N_{i,j} S_{ij}(M)Y_{ij}Y_{ji}} \sim \prod ^N_{i,j}
S^{-1/2}_{ij}(M)
\ee
(a constant factor is omitted again), and substitute the explicit
expression for $U_{ij}(M)$. If potential is represented as a formal series,
\be\label{d6}
V(Y) =\sum^\infty _{k=1}\frac{v_k}{k}Y^k
\ee
(and thus is supposed to be analytic in  $Y$  at  $Y = 0)$, the definition
(\ref{qv}) implies that
\be
S_2(M,Y)\,=\,\frac{1}{2}\,\sum_{k=2}^\infty v_k
\left\{\sum_{a+b=k-2}\mbox{Tr}\,M^aYM^bY\right\}
\ee
and, consequently,
\be
S_{ij} =\sum ^\infty _{k=2}v_k\Big\{
\sum _{a+b=k-2}\mu ^a_i\mu ^b_j \Big\}  =
\sum ^\infty _{n=0}V_k \frac{\mu^k_i-\mu^k_j}{\mu_i-\mu_j}=\\
 = \frac{V'(\mu _i)-V'(\mu _j)}{\mu_i-\mu _j}
\ee
Hence,
\be\label{d7}
\int e^{-S_2(M,Y)}dY\,=\,\frac{\Delta(\mu)}{\Delta(V'(\mu))}\,
\prod_{i=1}^N[V''(\mu_i)]^{-1/2}
\ee
and substitution of (\ref{d1}), (\ref{d4}) and (\ref{d7}) to (\ref{def})
gives the following representation of the GKM partition function:
\be\label{g3}
Z^V_{_N}[M]\,=\,\frac{\Delta(V'(\mu))}{\Delta(\mu)}
\left.\left.\prod_{i=1}^N\right\{[V''(\mu_i)\big]^{-1/2}
e^{V(\mu_i)-\mu_iV'(\mu_i)}\right\}
F[V'(M)]\,\equiv\\ \equiv\,
\frac{\det\,\Phi^V_i(\mu_j)|_{i,j=1}^N}{\Delta(\mu)}
\ee
where
\be\label{v1}
\Phi^V_i(\mu)\,=\,[V''(\mu)]^{1/2}\,e^{V(\mu)-\mu V'(\mu)}\,
\int x^{i-1}e^{-V(x)+xV'(\mu)}dx
\ee

\subsection{Functional relations}

The vectors (\ref{v1}) form a linear independent infinite set. In the
generic situation the basis vectors determining the $\tau$-function
are functionally independent since they are parametrized by arbitrary
$gl_\infty$  matrix. On the contrary, in GKM case the solution is parametrized
, loosely speaking, by the vector (the coefficients of the potential $V(x)$).
In this sense the solution (\ref{g3}) is degenerate; the degeneration results
to functional relations (the constraints) on the level of the basis vectors
which, in turn, can be considered as a definition of GKM from the Grassmannian
point of view \cite{SP}.\\
Consider the model, parametrized by arbitrary polynomial potential of degree
$p+1\,;\;\,p\geq 2$:
\be
V(x)\,=\,\sum_{k=1}^{p+1}\frac{v_k}{k}x^k
\ee
First of all, after multiplication of (\ref{v1}) by $V'(\mu)$,
the integration by parts gives (assuming the vanishing boundary conditions):
\be
V'(\mu)\Phi^V_i(\mu)\,=\,[V''(\mu)]^{1/2}\,e^{V(\mu)-\mu V'(\mu)}\,
\int x^{i-1}e^{-V(x)}\frac{\d}{\d x}e^{xV'(\mu)}dx\,=\\
=\,[V''(\mu)]^{1/2}\,e^{V(\mu)-\mu V'(\mu)}\,
\int \Big\{x^{i-1}V'(x)-(i-1)x^{i-2}\Big\}e^{-V(x)+xV'(\mu)}dx\
\ee
i.e.
\be\label{g7}
V'(\mu)\Phi^V_i(\mu)\,=\,\sum_{k=1}^{p+1}v_k\Phi^V_{i+k-1}(\mu)-
(i\!-\!1)\Phi^V_{i-1}(\mu)\;;\hspace{1cm}i=1, 2,\ldots
\ee
This relation generalizes the notion of $p$-reduced KP hierarchy;
for the monomial potential one gets the condition
(\ref{p-r}) exactly. We shall show below (Sect. \ref{eq-h}) that the general
constraint (\ref{g7}) has the natural interpretation in terms of equivalent
hierarchies.\\
There is another type of constraint which is a generalization of (\ref{vr1}).
Indeed,
\be\label{g4}
\Phi^V_i(\mu)\,=\,[V''(\mu)]^{1/2}\,e^{V(\mu)-\mu V'(\mu)}\,
\frac{1}{V''(\mu)}\frac{\d}{\d\mu}
\int e^{-V(x)+xV'(\mu)}dx\,\equiv\,A^V(\mu)\Phi^V_{i-1}(\mu)
\ee
where $A^V(\mu)$ is the first-order differential operator of a special
form
\be\label{g5}
A^V(\mu)\,=\,\frac{e^{V(\mu)-\mu V'(\mu)}}{[V''(\mu)]^{1/2}}\,
\frac{\d}{\d\mu}\,\frac{e^{-V(\mu)+\mu V'(\mu)}}{[V''(\mu)]^{1/2}}\,=\\
=\,\frac{1}{V''(\mu)}\frac{\d}{\d\mu}\,+\,\mu\,-\,
\frac{V'''(\mu)}{2[V''(\mu)]^2}
\ee
Thus, we have the functional relation
\be\label{g6}
\Phi^V_{i+1}(\mu)\,=\,A^V(\mu)\Phi^V_i(\mu)
\ee
which leads to a kind of string equation similar to (\ref{vr2}). To obtain
the differential (w.r.t time variables) constraint on GKM partition function
resulting from (\ref{g6}), the notion of the quasiclassical hierarchies is
required (Sect. \ref{qua}).

\subsection{GKM as a solution of KP hierarchy}
We proved that the GKM partition function (\ref{def}) is represented in
the determinant form
\be\label{g3'}
Z_N^V[M]\,=\,\frac{\det\,\Phi^V_i(\mu_j)|_{i,j=1}^N}{\Delta(\mu)}
\ee
where the vectors $\Phi^V_i(\mu)$ are defined by (\ref{v1}).
Moreover, using the steepest descent method it is not hard
to find the following asymptotics of the GKM basis vectors:
\be\label{v2}
\Phi^V_i(\mu)\,=\,\left.\left.\mu^{i-1}\right(1+O(\mu^{-p-1})\right)
\hspace{1cm}\mu\to\infty
\ee
- compare with (\ref{det2}), (\ref{as2}).
From above consideration it follows that, being written in Miwa times
(\ref{mi}), the partition function (\ref{g3'}) solves the KP hierarchy,
i.e.
\be
Z[T]\,\sim\,\tau_n(T)
\ee
with some (yet unknown) value of the vacuum state $n$ (see definition
(\ref{tauTL})).

\bigskip\noindent
Before proceed further, the important remark
concerning the dependence on $N$ in the formula (\ref{g3'}) deserve
mentioning. The entire set $\{\Phi^V_i(\mu )\}$ is certainly $N$-independent
and infinite. It is evident that $\Phi^V_i$'s are linear independent.
The r.h.s. of (\ref{g3'}) naturally represents the
$\tau$-function for an {\it infinitely large} matrix $M$. In order
to return to the case of finite $N$, it is enough to require that all
eigenvalues of  $M$, except  $\mu _1,\ldots,\mu _N$, tend to infinity. In this
sense the partition function  $Z_N^V[M]$ is
independent of $N$; the entire dependence on $N$ comes from the argument
$M$: $N$ is the quantity of finite eigenvalues of $M$. As a simple check
of consistency, let us additionally carry  $\mu _N$ to infinity in
(\ref{g3'}), then, according to (\ref{v2}),
\be
\mbox{det}_{_{\!N}}\Phi^V_i(\mu_j)=(\mu_{_N})^{N-1}\cdot \mbox{det}_{_{\!N-1}}
\Phi^V_i(\mu _j)\cdot(1 +O(1/\mu_{_N}))
\ee
and
\be
\Delta_{_{\!N}}(\mu)\sim(\mu_{_N})^{N-1}\Delta_{_{\!N-1}}(\mu)(1 +
O(1/\mu_{_N}))
\ee
Therefore,
\be
Z_N^V[M] \stackreb{\mu_{_N}\to\infty}{\sim} Z^V_{N-1}[M]\cdot(1+O(1/\mu_{_N}))
\ee
This is the exact statement about the $N$-dependence of the GKM partition
function. In this sense one can claim that GKM partition function is
independent on $N$.
Therefore, we often omit the subscript $N$ in what follows.

\bigskip\noindent
As the solution of the KP hierarchy, the partition function
(\ref{g3'}) is parametrized by the coefficients
of the polynomial $V$. Since the latter depends only on the finite number
of parameters, the original matrix integral describes very particular
$\tau$-function. Therefore, the question arises whether it possible to write
down some kind of constraints which naturally select this specific solution
from the huge set of the typical $\tau$-functions parametrized by $gl_\infty$
matrix $||A_{ij}||$ (\ref{point}). It turns out, that GKM $\tau$-function
satisfies the subset of $W_{1+\infty}$ constraints;
indeed, one can find a number of the differential (in KP times $\{T_k\}$)
operators which annihilate the function (\ref{g3'}). This gives the
invariant description of the model in the spirit of \cite{FKN1}. The problem
is to describe the action of these operators on the $\tau$-function
which is essentially written in the Miwa variables. Of course, due to
\cite{KS}, \cite{FKN2} it is well known how to reformulate all the
constraints on the level of the basis vectors: the complete information
concerning the invariant properties of the $\tau$-functions can be decipher
from the relations similar to (\ref{g6}), (\ref{g7}) and vice versa;
this has been demonstrated explicitely in sections \ref{td} and \ref{virs}.
In the case of monomial potential the invariant properties
of the basis vectors give, indeed, the complete information (see below).
It is important, however, that relations mentioned above are not
enough to describe the non-trivial evolution of the GKM partition
function w.r.t. deformations of the potential $V$ (say, from the monomial
to arbitrary polynomial of the same degree). The account of such
deformations results to highly involved mixture of the standard KP flows and
so-called quasiclassical (or dispersionless) ones.
In order to interprets the latter evolution one needs to know the action
of the operators which do not annihilate the $\tau$-function of GKM. The
non-invariant actions can not be reformulated in terms of the basis
vectors; the explicit formulae on the level of $\tau$-functions are required.

\subsection{GKM with monomial potential. $p$-reduced KP hierarchy and
$\mbox{L}_{-p}$ constraint}

Consider the GKM partition function in the simplest case of monomial
potential $V(X)=\frac{\textstyle X^{p+1}}{\textstyle p+1}$:
\be\label{r0}
Z^{(p)}[M]\,=\,\frac{
\raise-3pt\hbox{$e^{-\frac{p}{p+1}\Tr M^{p+1}}$}{\displaystyle\int}
\raise-3pt\hbox{$dX e^{\Tr\,\big[- \frac{X^{p+1}}{p+1}+M^pX\big]}$}}
{\displaystyle\int
\raise-2pt\hbox{
$dX e^{-\frac{1}{2}\Tr\,\big[\sum_{a+b=p-2}M^aXM^bX\big]}$}}
\,=\,\frac{\det \Phi^{(p)}_i(\mu_j)}{\Delta(\mu)}
\ee
The basis vectors
\be\label{r2}
\Phi^{(p)}_i(\mu)\,\equiv\,\sqrt{p\mu^{p-1}}\,e^{-\frac{\scriptstyle p}
{\scriptstyle p+1}{\mu^{p+1}}}
\int x^{i-1}\,e^{-\frac{\scriptstyle x^{p+1}}{\scriptstyle p+1}+x\mu^p}dx
\ee
satisfy the obvious relations
\be\label{r3}
\mu^p\,\Phi_i^{(p)}(\mu)\,=\,\Phi^{(p)}_{i+p}(\mu)\,-\,
(i-1)\Phi_{i-1}^{(p)}(\mu)
\ee
\be\label{r4}
A^{(p)}(\mu)\,\Phi_i^{(p)}(\mu)\,=\,\Phi_{i+1}^{(p)}(\mu)
\ee
where
\be\label{r5}
A^{(p)}(\mu)\,\equiv\,\frac{1}{p\mu^{p-1}}\frac{\d}{\d\mu}\,-\,
\frac{p-1}{2p\,\mu^p}+\,\mu
\ee
is the Kac-Schwarz operator \cite{KS}. Note that up to linear term
it is proportional to the Virasoro operator $A_{-p}$ defined in (\ref{Vir4}).

\bigskip\noindent
We have seen already that the partition function (\ref{r0}) is a
$\tau$-function of KP hierarchy. Now more concrete statements can be made.
First of all, the GKM $\tau$-function is a solution of $p$-reduced KP
hierarchy. Moreover, $Z^{(p)}[T]$ is independent of times $T_{np}$:
\be\label{r6}
\frac{\d Z^{(p)}[T]}{\d T_{np}}\,=\,0 \;,\hspace{1cm}n=1,2,\,\ldots
\ee
In addition, the partition function (\ref{r0}) satisfies the
$\L_{-p}$ constraint:
\be\label{r7}
\frac{1}{p}\,\L_{-p}Z^{(p)}[T]+\frac{\d Z^{(p)}[T]}{\d T_1}\,=\,0
\ee
Let us give some comments concerning the relation (\ref{r6}).
Due to (\ref{v2}) one should note that first $p+1$
vectors $\Phi^{(p)}_1(\mu),\,\ldots\,,\Phi^{(p)}_{p+1}(\mu)$ has a canonical
structure (\ref{as1}). Therefore, for $k=p$ the formula (\ref{3der}) holds if
one substitutes $\Phi^{(p)}_i\,,\;i=1,\,\ldots\,N$ instead of canonical GKM
vectors $\Phi^{(can)}_i$ (see more careful discussion of this point in
Sect. \ref{gen}). Moreover, due to (\ref{r3}) the combination
$\Phi^{(p)}_{i+p}-\mu^p\Phi^{(p)}_i$ does not contain the vector
$\Phi^{(p)}_i$. Hence, from (\ref{p-r2})
\be\label{r8}
\frac{\d Z^{(p)}[T]}{\d T_p}\,=\,0
\ee
From general KP theory one can deduce that constraint (\ref{r8}) implies all
higher relations of the form $\d_{T_{np}}Z^{(p)}[T]= Const\cdot\,
Z^{(p)}[T]$. Actually, it follows from the relations (\ref{r3}) due to
discussion in Sect. \ref{td}. Thus, $Z^{(p)}[T]$ is, indeed, the
$\tau$-function of $p$-reduced KP hierarchy, i.e. the corresponding Lax
operator satisfies the constraint
\be\label{lp}
L^p\,=\,(L^p)_{_+}
\ee
The {\it simple} proof of more strong statement (\ref{r6}), namely,
the complete independence of times $T_{np}\,,\;n\geq 1$ is absent,
unfortunately (see \cite{GKM} and, especially, \cite{IZ1} for details).

\bigskip\noindent
To derive the constraint (\ref{r7}) one needs again the canonical
structure of the GKM vectors
$\Phi^{(p)}_1(\mu),\,\ldots\,,\Phi^{(p)}_{p+1}(\mu)$.
It is important that because of this fact the
relations (\ref{time}) (with $k=1$) and (\ref{Vir4}) (with $k=p$) can be
written in terms of $\{\Phi^{(p)}_i\}$. The Kac-Schwarz operator (\ref{r5})
coincides with the formal shift operator $B(\mu)$ (\ref{sh}) due to
(\ref{r4}) while $A_{-p}(\mu)$ in (\ref{Vir4}) is represented as
$p(A^{(p)}(\mu)-\mu)$. Hence, one arrives to relations
\be\label{r10}
\frac{\d Z^{(p)}}{\d T_1}\,=\,\frac{1}{\Delta(\mu)}\sum_{m=1}^N
\left(A^{(p)}(\mu_m)-\mu_m\right)\det\Phi^{(p)}_i(\mu_j)
\ee
\be\label{r11}
\L_{-p}Z^{(p)}\,=\,
\,=\,-\,p\,\frac{1}{\Delta(\mu)}\,\sum_{m=1}^N\,
\left(A^{(p)}(\mu_m)-\mu_m\right)\,\det\Phi^{(p)}_i(\mu_j)
\ee
thus getting the constraint (\ref{r6}). Note that the latter can be written
in the form
\be\label{r12}
\frac{1}{2p}\sum_{k=1}^{p-1}k(p-k)T_kT_{p-k}\,+\,\frac{1}{p}
\sum_{k=1}^\infty(k+p)\Big(T_{k+p}+\frac{p}{p+1}\delta_{k,1}\Big)
\frac{\d \log Z^{(p)}}{\d T_k}\,=\,0
\ee
To conclude, the GKM $\tau$-function with monomial potential satisfies
the usual $\L_{-p}$-constraint (the integrated version of the string equation)
with the shifted times
\be\label{r13}
T_k\,\to\,T_k+\frac{p}{p+1}\delta_{k,p+1}
\ee

\subsection{General case: $V'$-reduction and transformation of times}
\label{gen}
In general situation of arbitrary polynomial of degree $p$
one gets the following matrix model:
\be\label{gr}
Z^V[T]\,=\,
\frac{e^{\raise2pt\hbox{$\scriptstyle \!\Tr\,[V(M)-MV'(M)]$}}}
{\int e^{-S_2(X,M)}dX}\,
\int e^{\raise2pt\hbox{$\scriptstyle \!\Tr\,[-V(X)+XV'(M)]$}}\;dX
\ee
The partition function (\ref{gr}) can be represented in the standard
determinant form:
\be\label{dv}
Z^V[T]\,=\,\frac{\det\Phi^V_i(\mu_j)}{\Delta(\mu)}
\ee
Therefore, $Z^V[T]$ is a $\tau$-function of KP hierarchy.
Its basis vectors
\be\label{gr0}
\Phi^V_i(\mu)\,=\,[V''(\mu)]^{1/2}\,e^{V(\mu)-\mu V'(\mu)}
\int x^{i-1}e^{-V(x)+xV'(\mu)}dx
\ee
satisfy the relations
\be\label{gr3}
V'(\mu)\Phi^V_i(\mu)\,=\,\sum_{k=1}^{p+1}v_k\Phi^V_{i+k-1}(\mu)-
(i\!-\!1)\Phi^V_{i-1}(\mu)\;;\hspace{1cm}i=1, 2,\ldots
\ee
\be\label{gr2}
\Phi^V_{i+1}(\mu)\,=\,A^V(\mu)\Phi^V_i(\mu)
\ee
where $A_{_V}(\mu)$ is the first-order differential operator
\be\label{gr1}
A^V(\mu)\,=\,\frac{1}{V''(\mu)}\frac{\d}{\d\mu}\,-\,
\frac{V'''(\mu)}{2[V''(\mu)]^2}+\mu
\ee
As before, these relations impose severe restrictions on the hierarchy.
It can be shown \cite{GKM} that $Z^V[T]$ satisfies the generalized
Virasoro constraint
\be\label{vg'}
\L^V\,Z^V[T]\,=\,0
\ee
where
\be\label{vg}
\L^V=\sum_{n\geq 1}\mbox{Tr}\,
\Big[\frac{1}{V''(M)M^{n+1}}\Big]\frac{\d}{\d T_k} -
{1\over 2}\sum _{i,j}\frac{1}{V''(\mu_i)V''(\mu_j)}
\frac{V''(\mu _i)\!-\!V''(\mu_j)}{\mu _i-\mu _j}+{\d\over\d T_1}
\ee
For monomial potential this constraint is reduced to (\ref{r12}) while, in
general, it is impossible to write the compact expression of (\ref{vg})
in original times (\ref{mi}). Nevertheless, one can construct the set of new
times $\{{\wt T}_k\}$ as a linear combinations of "old" ones, $\{T_k\}$, in
such a way that the operator (\ref{vg}) can be transformed to the standard
one being expressed in $\wt{T}_k$. The way to find appropriate linear
combinations is as follows. From (\ref{gr3}) one sees that GKM basis vectors
determine the invariant point of the Grassmannian such that
\be\label{gr4}
{\cal P}(\mu)\Phi^V_i(\mu)\subset\mbox{Span}\,\{\Phi^V(\mu)\}\,;
\hspace{1cm}{\cal P}(\mu)\equiv V'(\mu)
\ee
This condition is a natural generalization of the standard $p$-reduction
and is called the $V'$-reduction.
The general ideology \cite{SP} tells us that the pseudo-differential Lax
operator
\be\label{gr5}
L\,=\,\d+u_2\d^{-1}+u_3\d^{-2}+\ldots\\
L\Psi\,=\,\mu\Psi
\ee
corresponding to this point obeys the property
\be\label{gr6}
[{\cal P}(L)]_{_-}\,=\,0
\ee
i.e. $V'(L)$ is a differential operator of order $p$. Therefore, there
exists the Lax operator of KP hierarchy
\be\label{gr7}
\wt L\,=\,\d+\wt u_2\d^{-1}+\wt u_3\d^{-2}+\ldots\\
\wt L\Psi\,=\,\wt\mu\Psi
\ee
such that
\be\label{gr8}
\wt L^p\,=\,{\cal P}(L)
\ee
and, certainly, the relation between the spectral parameters of the
corresponding hierarchies is
\be\label{gr9}
\wt\mu\,=\,{\cal P}^{1/p}(\mu)
\ee
Now it is clear that the relevant spectral parameter is $\wt\mu$ rather
then $\mu$.
Therefore, the times appropriate for description of
$V'$-reduced KP hierarchy should be determined by relations
\be\label{gr10}
\wt{T}_k\,=\,-\,\frac{1}{k}\sum_i\wt\mu^{-k}_i\,\equiv\,
-\,\frac{1}{k}\sum_i{\cal P}^{-k/p}(\mu_i)
\ee
In order to find the relation between $\{T_k\}$ and $\{\wt{T}_k\}$
one introduces the notion of the residue operation $\mbox{Res}$. For any
Laurent series $F(\la)=\sum_k F_k\la^k$
\be\label{res}
\mbox{Res}\,F(\la)d\la\,=\,F_{-1}
\ee
It is easy to see that this operation satisfies the properties
\be\label{res0}
\mbox{Res}\,\frac{dF(\la)}{d\la}\,d\la\,=\,0\\
\mbox{Res}\,Fd_\la G\,=\,-\,\mbox{Res}\,Gd_\la F\\
\mbox{Res}\,Fd_\la G\,=\,\mbox{Res}\,F_{_+}d_\la G_{_-}\,+\,
\mbox{Res}\,F_{_-}d_\la G_{_+}
\ee
for any two Laurent series $F(\la)\equiv F_{_+}(\la)+F_{_-}(\la)$ and
$G(\la)\equiv G_{_+}(\la)+G_{_-}(\la)$ where $F_{_+}\,(F_{_-})$ are the
parts of the corresponding Laurent series containing only non-negative
(negative) powers in $\la$.\\
Using the properties of $\mbox{Res}$ one finds
relations:
\be\label{gr11}
\wt{T}_k\,=\,\frac{1}{k}\sum_{m=k}^\infty
mT_m\,\mbox{Res}\,\la^{m-1}{\cal P}^{-k/p}(\la)d\la
\ee
\be\label{gr12}
T_k\,=\,\sum_{m=k}^\infty
\wt{T}_m\,\mbox{Res}\,\la^{-k-1}{\cal P}^{m/p}(\la)d\la
\ee
Let us prove now that for arbitrary polynomial potential the GKM partition
function (\ref{gr}) is independent of time $\wt{T}_p$:
\be\label{gr13}
\frac{d Z^V[T(\wt{T})]}{\d \wt{T}_p}\,=\,0
\ee
i.e. $V'$-reduced KP hierarchy resembles the standard $p$-reduction while
considering the evolution along new integrable flows $\wt T_k$.

\bigskip\noindent
Actually, we shall derive more general formulas for the derivatives of $Z^V$
w.r.t first $p$ times $\wt T_1,\,\ldots\,,\wt T_p$.
To do so, one needs to calculate the derivatives w.r.t old times.
Let us apply the formula
(\ref{3der}) to the GKM partition function. Immediately the problem appears.
Indeed, the GKM vectors (\ref{gr0}) have a nice integral representation, but
these are not the canonical ones because of asymptotics (\ref{v2}). On the
other hand, the formula (\ref{3der}) is valid only for canonical basis
vectors. The compact integral representation for $\Phi^{(can)}_i(\mu)$ is
absent for GKM. Therefore, it is impossible to find the matrix integral
representation for the derivatives $\d_{_{T_k}}Z^V$
which is valid for {\it all} times $k\geq 1$.
Fortunately, this problem disappears while
considering the derivatives w.r.t. first $p$ times $T_1,\,\ldots\,T_p$. The
key point is that the first $p+1$ basis vectors
$\Phi^V_1(\mu),\ldots,\Phi^V_{p+1}(\mu)$ already have a canonical form
(see (\ref{v2})).
As a corollary, one can directly use (\ref{3der}) with the simple
substitution $\Phi^{(can)}\,\to\,\Phi^V(\mu)$ (i.e. without any modification)
for derivatives with respect to these times. The derivatives w.r.t. higher
times do not allow such replacements. To illustrate this statement,
one can check the "marginal" derivative $\d_{_{T_p}}Z^V$ which contains,
for example, the particular term (see (\ref{3der}) with $k=p$)
\be\label{2der}
\frac{1}{\Delta(\mu)}\,\left|
\begin{array}{ccc}
\Phi^{(can)}_1(\mu_1) &\ldots&
\Phi^{(can)}_1(\mu_N)\\
\ldots &\ldots&\ldots \\
\Phi^{(can)}_{N-1}(\mu_1) &\ldots&\Phi^{(can)}_{N-1}(\mu_N)\\
\Phi^{(can)}_{N+p}(\mu_1) &\ldots&\Phi^{(can)}_{N+p}(\mu_N)
\end{array}\right|
\ee
Obviously, the first $N-1$ rows in this expression can be written in
terms of the GKM vectors (\ref{gr0}). The only trouble can come from the last
row. But due to asymptotics (\ref{v2}) the canonical vectors can
be represented in the GKM basis as
$\Phi^{(can)}_{N+p}=\Phi^V_{N+p}+(\al_{_{N+p}}\Phi^V_{N-1}+\mbox{lower terms})$
with some constant $\al_{_{N+p}}$ and the row with entries
$\al_{_{N+p}}\Phi^V_{N-1}(\mu_1),\,\ldots\,, \al_{_{N+p}}\Phi^V_{N-1}(\mu_N)$
(as well as the rows with lower terms) does not contribute to determinant
(\ref{2der}). This conclusion is true for all other determinants resulting
to $\d_{_{T_p}}Z^V$. Hence, the formula (\ref{3der}) with $k=p$ remains
unchanged if one simply substitutes $\Phi^V_i$ instead of $\Phi^{(can)}_i$.
The same reasoning is applied, certainly, for all derivatives
$\d_{_{T_k}}Z^V$ with $k\leq p$.
On the contrary, the derivative $\d_{_{T_{p+1}}}Z$ contains the determinant
similar to (\ref{2der}) with
$\Phi^{(can)}_{N+p+1}(\mu_1),\,\ldots\,, \Phi^{(can)}_{N+p+1}(\mu_N)$
in the last row. In this case the transformation
$\Phi^{(can)}_{N+p+1}=\Phi^V_{N+p+1}+(\al_{_{N+p+1}}\Phi^V_{N}+\mbox{lower
terms})$
results to additional term proportional to $Z^V$. Evidently, the higher
derivatives become more and more involved while expressing them through
non-canonical vectors (\ref{gr0}).\\
Due to the reasons described above only the first $p$ derivatives have a
simple integral representations. In this case the formula (\ref{3der}) gives
for $1\leq k\leq p$
\be\label{t1}
\frac{\d Z^V[T]}{\d T_k}\,=\,
\frac{e^{\raise2pt\hbox{$\scriptstyle \!\Tr\,[V(M)-MV'(M)]$}}}
{\int e^{-S_2(X,M)}dX}\;\int \mbox{Tr}\,[X^k-M^k]
\,e^{\raise2pt\hbox{$\scriptstyle \!\Tr\,[-V(X)+XV'(M)]$}}\;dX
\ee
or, in compact notations,
\be\label{t2}
\frac{\d}{\d T_k}\log Z^V[T]\,=\,\<\mbox{Tr}\,X^k-\mbox{Tr}\,M^k\>\;;
\hspace{1cm}1\leq k\leq p
\ee
where
\be\label{aver}
\<{\cal F}(X)\>\,\equiv\,\frac{\int
{\cal F}(X)\,e^{\raise2pt\hbox{$\scriptstyle \!\Tr\,[-V(X)+XV'(M)]$}}\;dX}
{\int \,e^{\raise2pt\hbox{$\scriptstyle \!\Tr\,[-V(X)+XV'(M)]$}}\;dX}
\ee
Indeed, the calculation of $\<\mbox{Tr}\,X^k\>$ does not differ
in technical details from those
resulting to (\ref{g3}) and gives exactly the r.h.s. of (\ref{3der})
with $\Phi^{(can)}$ substituted by $\Phi^V$
\footnote{
The fermionic approach together with the above reasoning allows to write
more complicated derivatives quite explicitely.
Without proof we represent the formula
$$
\frac{\d^2 \log Z^V_N}{\d T_k\d T_m}=
\left\<\Big(\mbox{\rm Tr}X^k-\mbox{\rm Tr}M^k\Big)
\Big(\mbox{\rm Tr}X^m-\mbox{\rm Tr}M^m\Big)\right\>\;;
\hspace{0.5cm}1\leq k+m\leq p $$
The generalization is evident.
}
.
Using the relation (\ref{gr12}) between old and new times it
is easy to find the formulas we need:
\be\label{t3}
\frac{\d}{\d \wt
T_k}\log Z^V[T(\wt T)]\,=\, \left\<\mbox{Tr}\,[{\cal P}^{k/p}(X)]_{_+}
-\mbox{Tr}\,[{\cal P}^{k/p}(M)]_{_+}\right\>\;; \hspace{1cm}1\leq k\leq p
\ee
Note that in the r.h.s. of (\ref{t3}) is expressed through the eigenvalues
of the transformed matrix $\wt M$, i.e. $M$ should be substituted by the
solution of equation
\be\label{gr14}
{\cal P}(M)\,=\,\wt M^p
\ee The relation
(\ref{gr13}) can be readily proved now. Indeed,
\be \frac{\d}{\d \wt T_p}\log
Z^V[T(\wt T)]\,=\, \<\mbox{Tr}\,V'(X)-\mbox{Tr}\,V'(M)\>
\ee
and the r.h.s. vanishes since the expression under integral is a total
derivative. We proved
that, being written in times $\{\wt{T}_k\}$, the partition function
(\ref{gr}) has something to do with a solution of $p$-reduced KP hierarchy.
Therefore, it is natural to expect that the complicated Virasoro constraint
(\ref{vg'}), (\ref{vg}) can be simplified also being represented as a
differential operator w.r.t $\{\wt{T}_k\}$. This expectation is {\it almost}
true but slightly premature: the point is that the partition function
$Z^V[T(\wt T)]$ is not a $\tau$ function in general. Indeed, to express the
partition function (\ref{gr}) in new times (\ref{gr10}) means to substitute
the spectral parameters $\{\mu_i\}$ entering in (\ref{dv}) by the (formal)
solution of equation (\ref{gr9}).  Evidently, the transformation
\be
\mu\,=\,\wt{\mu}(1+O(\wt{\mu}^{-1}))
\ee
destroys the structure of the van der Monde determinant and, hence, the
function $Z^V[M(\wt M)]$ does not possesses the standard form.
Nevertheless, the situation can be repaired: one can extract the genuine
$\tau$-function of the $p$-reduced KP hierarchy from $Z^V[M(\wt M)]$.
To describe the procedure we need to elaborate the notion of the equivalent
hierarchies.

\section{Equivalent hierarchies}\label{eq-h}

\subsection{Definition}
Consider the spectral problem $L\Psi=\mu\Psi$ where the operator $L$
defining the KP hierarchy has a standard form (\ref{l-op}). For any
given function $f$
\be\label{f}
f(\mu)=\sum_{i=-\infty}^0 f_i\mu^{i+1}\hspace{1cm}f_0=1
\ee
with time independent coefficients one can construct the new $L$-operator
\be\label{eq1}
\wt{L}\,=\,f(L)
\ee
which has the same structure as the original one. The spectral problem now
is
\be
\wt{L}\Psi\,=\,\wt{\mu}\Psi
\ee
where
\be\label{tm}
\wt{\mu}\,\equiv\,f(\mu)
\ee
The new operator $\wt L$ (\ref{eq1}) determines the KP hierarchy which is
called the equivalent to the original one \cite{S}. Introducing the
differential operators $\wt{B}_k\equiv (\wt{L}^k)_{\!{_+}}$, one can
construct the evolution equations
\be\label{eq2}
\frac{d\wt{L}}{\d\wt{T}_k}\,=\,[\wt{B}_k,\wt{L}]
\ee
which can be considered as a definition of times $\{\wt{T}_i\}$. Obviously,
\be
\wt{B}_m\,=\,B_k\,\frac{\d T_k}{\d \wt{T}_m}
\ee
The question is what is the relation between the solutions of the equivalent
hierarchies determined by the operators $L$ and $\wt{L}$. First of all, one
needs to establish the explicit relationship between $\{T_i\}$ and
$\{\wt{T}_i\}$. The second step is to find the $\tau$-function of the
"deformed" $\wt{L})$-hierarchy  which corresponds to arbitrary given function
$\tau(T)$ of the original $L$-hierarchy. This gives the precise mapping
between the equivalent hierarchies.

\subsection{Variation of the spectral parameter}
It is evident that relation (\ref{tm}) can be considered as a transformation
of the original spectral parameter $\mu$ under the action of the Virasoro
generators. Let
\be\label{Vsum} \sum_{k=1}^\infty
a_kA_{-k}(\mu)\,\equiv\,\frac{1}{W'(\mu)}\frac{\d}{\d\mu}+
\frac{1}{2}\,\Big(\frac{1}{W'(\mu)}\Big)'\;\equiv\;A(\mu)
\ee
where the differential operators $A_k(\mu)$ are determined in (\ref{vir2}).
The function $W(\mu)$ has the asymptotical behavior
\be\label{as3}
W'(\mu)=\frac{\mu^{s-1}}{a_s}\Big(1+O(\mu^{-1})\Big)\hspace{1cm}
\mu\,\to\,\infty
\ee
where $a_s$ is a first non-zero coefficient in the sum (\ref{Vsum}).
The exponential operator $\exp A(\mu)$ can be disentangled as
\be\label{dis1}
\exp\Big\{\frac{1}{W'(\mu)}\frac{\d}{\d\mu}+
\frac{1}{2}\,\Big(\frac{1}{W'(\mu)}\Big)'\Big\}\;=\;
\Big\{\d_\mu\Big(W^{-1}(W(\mu)+1)\Big)\Big\}^{1/2}\,
\exp\Big(\frac{1}{W'(\mu)}\frac{\d}{\d\mu}\Big)
\ee
where $W^{-1}$ is the function inverse to $W$. It is convenient to introduce
the function
\be\label{fw}
f(\mu)\,=\,W^{-1}\Big(W(\mu)+1\Big)\,\equiv\,\wt{\mu}
\ee
which has the Laurent expansion
\be
f(\mu)=\mu\Big(1+O(\mu^{-1})\Big)\hspace{1cm}\mu\,\to\,\infty
\ee
due to (\ref{as3}). We describe the transformation of the spectral parameter
by the formula
\be\label{mu-tr}
e^{{\textstyle\frac{1}{W'(\mu)}\frac{\d}{\d\mu}}}
\,\mu\,=\,W^{-1}\Big(W(\mu)+1\Big)\,\equiv\,f(\mu)
\ee
We have seen that the action of the operator $e^{A(\mu)}$ is expressed
in terms of the function $f$ rather than $W$. Therefore,
we shall denote below the function $A(\mu)$ entering in the definition
(\ref{Vsum}) as $A_f(\mu)$ keeping in mind the relation (\ref{fw})
between functions $f(\mu)$ and  $W(\mu)$. Note that the relations between
the coefficients $a_k$ and $f_i$ are rather complicated.

\bigskip\noindent
Introduce two sets of times
\be
T_k\,=\,-\,\frac{1}{k}\sum_i\mu_i^{-k}\;,\hspace{0,5cm}
\wt{T}_k\,=\,-\,\frac{1}{k}\sum_i\wt{\mu}_i^{-k}
\ee
where $\wt{\mu}=f(\mu)$. It is easy to find the relations between
these times using the residue operation:
\be\label{eq3}
\wt{T}_k\,=\,\frac{1}{k}\sum_{m=k}^\infty
mT_m\,\mbox{Res}\,\la^{m-1}f^{-k}(\la)d\la
\ee
\be\label{eq4}
T_k\,=\,\sum_{m=k}^\infty
\wt{T}_m\,\mbox{Res}\,\la^{-k-1}f^m(\la)d\la
\ee
in complete analogy with (\ref{gr11}), (\ref{gr12}) where
$f(\la)={\cal P}^{1/p}(\la)$. Note that in the operator form
\be\label{TT1}
\wt{T}_k(\{\wt{\mu}\}) \,\equiv\,
\,-\,\frac{1}{k}\sum_{i}\wt{\mu}_i^{-k}\,=\,
\prod_ie^{{\textstyle\frac{1}{W'(\mu_i)}\frac{\d}{\d\mu_i}}}\,T_k(\{\mu\})
\ee
Since
\be
W'(\wt{\mu})d\wt{\mu}\,=\,W'(\mu)d\mu
\ee
the transformation (\ref{eq4}) can be written as
\be\label{TT2}
T_k(\{\mu\}) \,\equiv\,
\prod_ie^{{\textstyle -\frac{1}{W'(\tilde{\mu}_i)}
\frac{\d}{\d\tilde{\mu}_i}}}\,\wt{T}_k(\{\wt{\mu}\})\,=\,
-\,\frac{1}{k}\sum_{i}\Big(f^{-1}(\wt{\mu}_i)\Big)^{-k}
\ee
where $f^{-1}$ is an inverse function to $f$
\footnote{
Note that
$$
f^{-1}(\wt{\mu})\,=\,W^{-1}\Big(W(\wt{\mu})-1\Big)
$$
to compare with (\ref{fw}).
}.

\subsection{$\tau$-functions of the equivalent hierarchies}

Consider the correspondence between the $\tau\!$-functions of the equivalent
hierarchies. Let $\tau(T)$ be a solution of the original hierarchy. Using the
relations (\ref{eq4}) one can consider it as a function of times
$\{\wt{T}_k\}$, i.e. to deal with $\tau[T(\wt{T})]$. It is reasonable to
assume that the latter object has something to do with the equivalent
hierarchy determined by the operator $\wt{L}$ (\ref{eq1}). Actually,
$\tau[T(\wt{T})]$ is not a solution of the equivalent hierarchy.
Nevertheless, being corrected by the appropriate factor, this function
do determines the solution we need. Namely, one can show that the expression
\be\label{rel0}
e^{^{\frac{1}{2}\,A_{km}\wt{T}_k \wt{T}_m}}\tau[T(\wt{T})]\,\equiv\,
\wt\tau(\wt{T})
\ee
with some definite matrix $A_{km}$ is a $\tau\!$-function of the equivalent
hierarchy. The most easiest way to prove this statement is to consider the
$\tau$-functions in the determinant form (\ref{det2}). It is clear that the
transformation of times $T\to \wt{T}$ corresponds to the transformation
of the Miwa variables $\mu_i=f^{-1}(\wt{\mu}_i)$. In terms of
$\wt{\mu}_i$ the original function $\tau[T(\wt{T})]$ is not the ratio of two
determinants (since
$\Delta(\mu)|_{\mu=f^{-1}(\wt{\mu})}\equiv\Delta(\mu(\wt\mu))$ is
not the Van der Monde determinant in terms of $\{\wt{\mu}_i\}$) and,
therefore, does not correspond to any $\tau$-function. Nevertheless,
the $\tau$-function of the equivalent hierarchy can be easily extracted.
Indeed, consider the identical transformation
\footnote{
By $f'(\mu(\wt\mu))$ we mean the function $\d_\mu f(\mu)$
calculated at the point $\mu=f^{-1}(\wt\mu)$.}:
\be\label{rel1}
\tau[T(\wt{T})]\,\equiv\,\Big\{\frac{\Delta(\wt{\mu})}{\Delta(\mu(\wt\mu))}\,
\prod_i [f'(\mu(\wt{\mu}))]^{1/2}\Big\}\,\wt\tau(\wt T)
\ee
where $\wt\tau(\wt T)$ as a function of times (\ref{TT1}) has the determinant
form  (\ref{det2}), i.e.
\be\label{det3}
\wt\tau(\wt T)\,=\,\frac{\det \wt\phi_i(\wt{\mu}_j)}{\Delta(\wt\mu)}
\ee
with the basis vectors
\be\label{ba1}
\wt\phi_i(\wt\mu)\,=\,[f'(\mu(\wt\mu))]^{-1/2}\,\phi_i(\mu(\wt\mu))
\ee
By the direct calculation one can show that the prefactor in the r.h.s.
of (\ref{rel1}) may be represented in the form
\be\label{id1}
\frac{\Delta(\wt{\mu})}{\Delta(\mu(\wt\mu))}\,\prod_i [f'(\mu(\wt{\mu}))]^{1/2}
\,=\,e^{-\frac{1}{2}\,A_{km}\wt{T}_k \wt{T}_m}
\ee
where
\be\label{a}
A_{km}\,=\,\mbox{Res}\,f^k(\la)d_\la (f^m(\la))_{_+}
\ee
Thus, one arrives to relation (\ref{rel0}).
We omit brute force derivation of the identity (\ref{id1})
because the technical details are not instructive here and represent below
the "physical" proof of (\ref{rel0}) on the level of the fermionic
correlators. Such approach has two advantages: first of all, it explains
very clearly the meaning of the identical redefinition in
(\ref{rel1})-(\ref{ba1}) and, besides, it describes the explicit
transformation of the point of the Grassmannian while passing to equivalent
hierarchy.

\subsection{Proof of equivalence formula}
Let us consider the identity
\be
\tau(T|g)\,\equiv\,\prod_ie^{{\textstyle -\frac{1}{W'(\tilde{\mu}_i)}
\frac{\d}{\d\tilde{\mu}_i}}}\, \,\tau(\wt{T}|g)
\ee
(see (\ref{TT2})). The last (rather trivial) relation can be reformulated
in terms of the fermionic correlators as follows.
Using the correspondence (\ref{start2}) between the fermionic correlators and
the $\tau$-functions one can write
\footnote{\label{fo} \label{1/2}
From (\ref{vir2}), (\ref{Vsum}), (\ref{dis1})-(\ref{mu-tr}) it is clear that
$$
e^{\vir_f(J)}\psi(\mu)\,e^{-\vir_f(J)}\,=\,(\d_\mu f(\mu))^{1/2}\psi(f(\mu))
$$
i.e. the fermions are transformed as 1/2-differentials.
The inverse transformation is
$$
e^{-\vir_f(J)}\psi(\wt\mu)\,e^{\vir_f(J)}\,=\,(\d_{\wt\mu}
f^{-1}(\wt\mu))^{1/2}\psi(f^{-1}(\wt\mu))\,\sim\,
e^{\textstyle -\frac{1}{W'(\tilde{\mu}_i)}\frac{\d}{\d\tilde{\mu}_i}}
\psi(\wt\mu)
$$
During the calculations in (\ref{qq}) we are using last of these two
formulas. In fact, the only things we need is that
$\psi(\mu)\sim e^{-\vir_f(J)}\psi(\wt\mu)\,e^{\vir_f(J)}$ and $\<N|\L_f=0$.}
\be\label{qq}
\<0|e^{H(T)}g|0\>\,=\,
\frac{\<N|\psi(\mu_N)\ldots\psi(\mu_1)g|0\>}
{\<N|\psi(\mu_N)\ldots\psi(\mu_1)|0\>}
\,=\,
\frac{\prod_ie^{{\textstyle
-\frac{1}{W'(\tilde{\mu}_i)}\frac{\d}{\d\tilde{\mu}_i}}}\,
\<N|\psi(\wt{\mu}_N)\ldots\psi(\wt{\mu}_1)g|0\>}
{\prod_ie^{{\textstyle -
\frac{1}{W'(\tilde{\mu}_i)}\frac{\d}{\d\tilde{\mu}_i}}}\,
\<N|\psi(\wt{\mu}_N)\ldots\psi(\wt{\mu}_1)|0\>} \equiv\\
\equiv\;
\frac{\,\<N|\psi(\wt{\mu}_N)\ldots\psi(\wt{\mu}_1)
e^{\vir_f(J)}g|0\>}
{\,\<N|\psi(\wt{\mu}_N)\ldots\psi(\wt{\mu}_1)
e^{\vir_f(J)}|0\>}\;=\;
\frac{\,\<0|e^{H(\wt{T})}e^{\vir_f(J)}g|0\>}
{\,\<0|e^{H(\wt{T})}e^{\vir_f(J)}|0\>}
\ee
where
\be\label{Vsum2}
\L_f(J)\,=\,\sum_{k=1}^\infty a_k\L_{-k}(J)
\ee
similarly to (\ref{Vsum}). Thus, we proved, that the transition to
the equivalent hierarchy results to the following identity between
$\tau$-functions of the corresponding hierarchies:
\be\label{rel}
\tau(T|g)\,=\,\frac{\tau(\widetilde T|e^{\mbox{\footnotesize\bf L}_f(J)}g)}
{\tau(\widetilde T|e^{\vir_f(J)})}
\ee
i.e. one needs to redefine the times together with the appropriate change
of the point of the Grassmannian
\be
g\;\to\;e^{\vir_f(J)}g\;\equiv\;g_{_f}
\ee
and perform simultaneously the renormalization of the $\tau$-function.
The formula (\ref{rel}) coincides with (\ref{rel1}).
Indeed, the numerator is a $\tau$-function which can be written in the
determinant form (\ref{det3}) with the basis vectors
\be
\wt{\phi}_i(\wt{\mu})\,=\,
\frac{\<0|\psi^\ast_{i-1}\psi(\wt{\mu})e^{\vir_f(J)}g|0\>}{\<0|g|0\>}
\hspace{1cm}i=1, 2, \ldots
\ee
These vectors coincide with the previously defined ones (\ref{ba1}). To
show this, let us move the Virasoro element to the left state $\<0|$. One
can discard the adjoint action of $e^{\vir_f}$ on $\psi^\ast_i$, since
\be
[\L_{-k}(J),\psi^\ast_{i}]\,=\,\Big(\frac{k-1}{2} -i\Big)\psi^\ast_{i-k}
\ee
and, by definition, $\L_f$ contains only the Virasoro generators
$\L_{-k}$ with $k>0$. Therefore, for any $i\ge 1$
\be
e^{\vir_f(J)}\psi^\ast_{i-1}e^{-\vir_f(J)}\,=\,\psi^\ast_{i-1}+
\mbox{lower modes}
\ee
But the negative modes annihilate the left state $\<0|$ while the
positive lower modes generate the lower basis vectors which do not contribute
to the determinant $\det \wt\phi_i(\wt\mu_j)$. Therefore, without
loss of generality one may assume
\be
\wt{\phi}_i(\wt{\mu})\,=\,
\frac{\<0|\psi^\ast_{i-1}e^{-\vir_f(J)}\psi(\wt{\mu})
e^{\vir_f(J)}g|0\>}{\<0|g|0\>}
\,=\,\Big(\d_{\widetilde\mu}f^{-1}(\widetilde\mu)\Big)^{1/2}
\;\frac{\<0|\psi^\ast_{i-1}
e^{{\textstyle -\frac{1}{W'(\tilde{\mu}_i)}\frac{\d}{\d\tilde{\mu}_i}}}\,
\psi(\wt{\mu})g|0\>}{\<0|g|0\>}\;\equiv\\
\equiv\;\Big(\d_\mu f(\mu(\widetilde\mu))\Big)^{-1/2}
\phi_i(\mu(\wt{\mu}))
\ee
where in the last equality the original basis vectors $\phi_i(\mu)$
are expressed in terms of the deformed spectral parameter $\wt{\mu}$
via the substitution $\mu\,=\,f^{-1}(\wt{\mu})$. We get the
basis vectors (\ref{ba1}). Note that the appearance of the normalization
factor in the definition (\ref{ba1}) is very natural due to the fact that the
basis vectors are transformed as 1/2-differentials under the
action of the Virasoro group (see footnote \ref{fo} on page \pageref{qq}).

\bigskip\noindent
Thus, the only problem is to calculate the trivial $\tau$-function
\be
\tau(\widetilde T|e^{\vir_f})\,=\,
\<0|e^{H(\wt{T})}e^{\vir_f(J)}|0\>\;=\;e^{\vir_f(T)}\cdot 1
\ee
which corresponds to the point of Grassmannian
\be\label{triv}
g_0\,=\,e^{\vir_f(J)}
\ee
It is evident that this function is the exponential of quadratic
combinations of $\wt{T}$. To find the explicit expression let us consider its
derivative with respect to the arbitrary time $\wt{T}_k$:
\be\label{der}
\d_{_{\wt{T}_k}}\tau(\widetilde T|e^{\vir_f})\,=\,
\<0|e^{H(\wt{T})}J_ke^{\vir_f(J)}|0\>
\ee
To calculate the r.h.s. of (\ref{der}) one can use the following trick.
The moving of the current $J_k$ through $e^{\vir_f(J)}$ to the right
state $|0\>$ results to arising of all lower modes $J_i$ with $i\leq k$ since
\be\label{LJ}
[\,\L_{-i},J_k]\,=\,-\,kJ_{k-i}
\ee
The positive
modes annihilate the right state and do not contribute to (\ref{der}). Then
move the negative modes through $e^{\vir_f(J)}$ back to the left. In such
commutation no positive modes arise (due to (\ref{LJ}) again). The
commutation of negative modes with $e^H$ gives the linear combination of
times due to commutation relations $[J_m,J_n]=m\delta_{n+m,0}$. At last, all
negative modes annihilate the left state $\<0|$ and the final result is
$\tau(\widetilde T|e^{\vir_f})$ multiplied by the linear combination of times.
The explicit calculation is as follows. From (\ref{LJ})
\be
[\,\L_f,J(\mu)]\,=\,\d_\mu\Big(\frac{1}{W'(\mu)}\,J(\mu)\Big)
\ee
and, therefore, the exponentiation gives the differential operator
$\exp\Big(\frac{1}{W'(\mu)}\d_\mu+(\frac{1}{W'(\mu)})'\Big)$ which can be
disentangled similarly to  (\ref{dis1}). Thus,
\be\label{j1}
e^{\vir_f}J(\mu)e^{-\vir_f}\,=\,\d_\mu f(\mu)\,J(f(\mu))
\ee
The inverse transformation is
\be\label{j2}
e^{-\vir_f}J(\mu)e^{\vir_f}\,=\,\d_\mu f^{-1}(\mu)\,J(f^{-1}(\mu))
\ee
Multiplying (\ref{j2}) by $\mu^k$ and taking the residue one gets
\be\label{j3}
J_ke^{\vir_f}\,=\,e^{\vir_f}\mbox{Res}\, f^k(\la)J(\la)d\la
\ee
and the action of this operator on the right state reduces to
\be\label{j4}
J_ke^{\vir_f}|0\>\,=\,e^{\vir_f}\mbox{Res}\, f^k(\la)J^{(-)}(\la)d\la|0\>
\ee
where $J^{(-)}(\la)$ denotes the linear combination of the negative current
modes in the expansion
\be
J(\la) \equiv \sum_{k=0}^\infty J_k\la^{-k-1}+\sum_{k=-\infty}^{-1}
J_k\la^{-k-1} \,\equiv\,J^{(+)}(\la)+J^{(-)}(\la)
\ee
Note that $J^{(-)}(\la)$ contains only non-negative degrees of the spectral
parameter $\la$. Recall that we denote
$(F(\la))_{_+}$ the part of the Laurent series $F(\la)$ containing
non-negative degrees of $\la$. From (\ref{j1})
\be\label{j5}
e^{\vir_f}J^{(-)}(\la)\,=\,
\Big(\d_\la f(\la)\,J^{(-)}(f(\la))\Big)_{\!+}e^{\vir_f}
\ee
One should stress that the positive modes $J^{(+)}(f(\la))$ do not
contribute to the r.h.s.  of the last formula since $\d_\la
f(\la)=1+O(\la^{-1})$ and $J^{(+)}(f(\la)$ contain only the negative degrees
of $\la$.  Combining (\ref{j4}) and (\ref{j5}) one gets
\be\label{j6}
J_ke^{\vir_f}|0\>\,=\,\mbox{Res}\, f^k(\la)
\Big(\d_\la f(\la)\,J^{(-)}(f(\la))\Big)_{\!+}d\la\;e^{\vir_f}|0\>\;\equiv\\
\equiv\;\sum_{m=1}^\infty\frac{1}{m}\,\mbox{Res}\,f^k(\la)
d_\la (f^m(\la))_{_+}
\,J_{-m}e^{\vir_f}|0\>
\ee
After substitution of (\ref{j6}) to (\ref{der}) and taking into account
the commutation relations
\be
e^{H(\wt{T})}J_{-m}e^{-H(\wt{T})}\,=\,J_{-m}+m\wt{T}_m
\ee
one arrives to equation
\be\label{der2}
\d_{_{\wt{T}_k}}\tau(\widetilde T|e^{\vir_f})\,=\,
\tau(\widetilde T|e^{\vir_f})\,\sum_{m=1}^\infty\,A_{km}\wt{T}_m
\ee
where
\be
A_{km}\,=\,\mbox{Res}\,f^k(\la)d_\la (f^m(\la))_{_+}
\ee
It is easy to show that $A_{km}=A_{mk}$. From (\ref{der2}) the final
answer is
\be
\<0|e^{H(\wt{T})}e^{\vir_f}|0\>\,=\,\exp\Big\{\frac{1}{2}
\sum_{k,m=1}^\infty\,A_{km}\wt{T}_k \wt{T}_m\Big\}
\ee
and the relation (\ref{rel}) between the $\tau$-functions of the equivalent
hierarchies takes the form
\be\label{taueq}
\tau[T(\wt T)|g]\,=\,
e^{-\frac{1}{2}\,A_{km}\wt{T}_k \wt{T}_m}\,
\tau(\wt{T}|e^{\vir_f}g)
\ee

\subsection{Equivalent GKM}
In the GKM context the equivalent hierarchies are naturally
described by the function
\be\label{s}
\wt\mu={\cal P}^{1/p}(\mu)\hspace{1cm}{\cal P}(\mu)\equiv V'(\mu)
\ee
Applying the general formula (\ref{rel1}) one can
represent the original partition function (written in new times $\wt{T}$)
as follows
\be\label{s0}
Z^V[T(\wt{T})]\,=\,\frac{\Delta(\wt\mu)}{\Delta(\mu)}\,
\prod_{i}\left(\frac{V''(\mu_i)}{p\,\wt\mu_i^{p-1}}\right)^{1/2}
\wt Z^V[\wt T]
\,=\,e^{-\frac{1}{2}\,A_{ij}\wt{T}_i \wt{T}_j}\,\wt Z^V[\wt T]
\ee
where
\be\label{s1}
A_{ij}\,=\,\mbox{Res}\,{\cal P}^{i/p}(\la)d_\la ({\cal P}^{j/p}(\la))_{_+}
\ee
and the $\tau$-function of the equivalent hierarchy is described by the
matrix integral
\be\label{s2}
\wt Z^V[\wt T]\,=\,\frac{\Delta(\wt{\mu}^p)}{\Delta(\wt\mu)}
\prod_i(p\,\wt\mu^{p-1}_i)^{1/2}\;\;
e^{\raise2pt\hbox{$\scriptstyle \!\Tr\,[V(M)-M\wt{M}^p]$}}
\int\,e^{\raise2pt\hbox{$\scriptstyle \!\Tr\,[-V(X)+X\wt{M}^p\,]$}}\;dX
\ee
The r.h.s. of (\ref{s2}) should be expressed in terms of the
matrix $\wt M$. Using the relation (\ref{s}) it is easy to find that
\be\label{s3'}
\mu\,=\,\frac{1}{p}\sum_{k=-\infty}^{p+1}kt_k\,\wt{\mu}^{k-p}
\ee
\be\label{s3}
V(\mu)-\mu V'(\mu)\,=\,-\sum_{k=-\infty}^{p+1}t_k\,\wt\mu^{k}
\ee
where
\be\label{s4}
t_k\,\equiv\,-\,\frac{p}{k(p\!-\!k)}\,
\mbox{Res}\,{\cal P}^{\frac{\scriptstyle p-k}{\scriptstyle p}}(\la)\,d\la
\ee
Note that the parameters $t_1,\,\ldots\, t_{p+1}$ are independent; they
are related with the coefficients of the potential $V$
and can be interpreted as the times generating some integrable
evolution. Indeed, (\ref{s4}) can be considered as a set of equations which
determine the coefficients of the potential as the functions of these
additional times.
The equations (\ref{s4}) naturally arise in the dispersionless KP hierarchy
(see below). Note also that all higher positive times are zero due to
polynomiality of ${\cal P}$ while the negative "times" $\{t_k\,,\;k<0\}$
are complicated functions of the independent positive times.\\
Substitution od (\ref{s3}) to (\ref{s2}) results to the
matrix integral which depends on two sets of times:
\be\label{s5}
\wt Z^V[\wt T,t]\,=\,
e\raise9pt\hbox{${\scriptscriptstyle\sum_{k=1}^\infty} {\scriptstyle
kt_{-k}\wt{T}_k}$} \left\{ \frac{\Delta(\wt{\mu}^p)}{\Delta(\wt\mu)}
\prod_i(p\,\wt\mu^{p-1}_i)^{1/2}\;\;
e\raise9pt\hbox{${\scriptscriptstyle -\sum_{k=1}^{p+1}}
{\;\scriptstyle t_{k}\Tr\,\wt{M}^k}$}\int\,
e^{\raise2pt\hbox{$\scriptstyle\!\Tr\,[-V(X)+X\wt{M}^p\,]$}}\;dX\right\}
\ee
where the coefficients of $V(X)$ are functions of quasiclassical times
$t_1,\,\ldots\,t_{p+1}$ according to (\ref{s4}). From (\ref{s1}) it follows
that $A_{i,np}=A_{np,i}=0$. Moreover, $t_{-p}=0$ and, due to (\ref{gr13}),
\be\label{s6}
\frac{\d \wt{Z}^V[\wt{T}]}{\d \wt{T}_p}\,=\,0
\ee

\bigskip\noindent
Thus, we extract the $\tau$-function of $p$-reduced KP hierarchy from the
general matrix integral (\ref{gr}). The last logical step to reveal the
genuine integrable object hidden in (\ref{gr}) is to consider the part
of partition function (\ref{s5}) without the exponential prefactor with a
linear $\wt{T}$-dependence, i.e. the matrix integral
\be\label{s7}
\tau^V[\wt T,t]\,=\,\frac{\Delta(\wt{\mu}^p)}{\Delta(\wt\mu)}
\prod_i(p\,\wt\mu^{p-1}_i)^{1/2}\;\;
e\raise9pt\hbox{${\scriptscriptstyle -\sum_{k=1}^{p+1}}
{\;\scriptstyle t_{k}\Tr\,\wt{M}^k}$}
\int\,e^{\raise2pt\hbox{$\scriptstyle \!\Tr\,[-V(X)+X\wt{M}^p\,]$}}\;dX
\ee
This is exactly the object we need. First of all,
this $\tau$-function has the standard determinant form
\be\label{s8}
\tau^V[\wt T,t]\,=\,\frac{\det \phi_i^V(\wt{\mu}_j)}{\Delta(\wt{\mu})}
\ee
with the basis vectors
\be\label{s9}
\phi^V_i(\wt{\mu})\,=\,\sqrt{p\,\wt{\mu}^{p-1}}\;
e\raise7pt\hbox{${\scriptscriptstyle -\sum_{k=1}^{\infty}}
{\scriptstyle t_{k}\,\wt{\mu}^k}$}
\int x^{i-1}e^{-V(x)+\wt{\mu}^px}dx
\ee
satisfying to $p$-reduction condition
\be\label{s10}
\wt{\mu}^p\phi^V_i\,=\,\sum_{j=1}^{p+1}v_j \phi^V_{i+j-1}-(i-1)\phi^V_{i-1}
\ee
as well as to Virasoro-type constraint:
\be\label{s11}
A(\wt{\mu})\phi^V_i\,=\,\phi^V_{i+1}
\ee
where
\be\label{s12}
A(\wt{\mu})\equiv\frac{1}{p\wt{\mu}^{p-1}}
\frac{\d}{\d\wt{\mu}}-\frac{p\!-\!1}{2p\,\wt{\mu}^p}
+\frac{1}{p}\sum_{k=1}^{p+1}kt_k\,\wt{\mu}^{k-p}
\ee
The partition function (\ref{s7}) possesses the remarkable
properties. First of all, it is a solution of the $p$-reduced KP hierarchy,
i.e.
\be
\frac{\d \tau^V[\wt{T},t]}{\d \wt{T}_p}\,=\,0
\ee
- this is the corollary of (\ref{s5}).
Further, the relation (\ref{s11}) implies that (\ref{s7})
satisfies the standard $\L_{-p}$-constraint
\be\label{s15}
\L^V_{-p}\tau^V[\wt{T},t]\,=\,0
\ee
\be\label{s16}
\L^V_{-p}=\frac{1}{2p}
\sum_{k=1}^{p-1}k(p-k)(\wt{T}_k\!+\!t_k)(\wt{T}_{p-k}\!+\!t_{p-k})
+\frac{1}{p}\sum_{k=1}^\infty(k\!+\!p)(\wt{T}_{k+p}\!+\!t_{p+k})
\frac{\d}{\d \wt{T}_k}
\ee
where the KP times are naturally shifted by the corresponding quasiclassical
ones. We shall give the direct proof of this statement in Sect. \ref{pvir}.

\bigskip\noindent
Moreover, the form of $\L_{-p}$-operator (\ref{s16}) give a hint that there
should exist the object depending only on the sum $T_k+t_k$. Indeed, this
is the case. Consider the product
\be\label{s13}
{\cal Z}^V[\wt{T},t]\,\equiv\,\tau^V[\wt{T},t]\,\tau_0(t)
\ee
where $\tau_0(t)$ is a $\tau$-functions of {\it quasiclassical $p$-reduced
KP hierarchy}. We show in Sect \ref{ptimes} that ${\cal Z}^V[\wt{T},t]$
depends only on the sum of KP and quasiclassical times:
\be\label{s14}
\left(\frac{\d}{\d \wt{T}_k}\,-\,
\frac{\d}{\d t_k}\right){\cal Z}^V[\wt{T},t]\,=\,0\;,\hspace{0.7cm}
k=1,\,\ldots\,p
\ee
Of course, ${\cal Z}^V$ also satisfies the constraint (\ref{s15}).\

\bigskip\noindent
To prove the above statements, some essentials concerning the quasiclassical
hierarchies are required. At this stage one sees that the GKM
includes almost all fundamental notions of the integrable theory
thus boiled these ingredients together.

\section{Quasiclassical KP hierarchy}\label{qua}
\subsection{Basis definitions}
The general treatment of the quasiclassical limit in the theory of the
integrable systems can be found in \cite{Kri1}-\cite {Du} and references
therein. Here we outline the description of so-called quasiclassical, or
dispersionless KP hierarchy \cite{TT} which is
appropriate limit of the standard KP hierarchy (the careful investigation of
this limit was given in \cite {TT2}). Consider the quasiclassical version of
the $L$-operator \footnote{In what follows we shall use the term "operator"
in order to keep the resemblance with the usual KP-terminology; but one
should perceive, certainly, that we are dealing with the functions, not with
the genuine operators.}
\be
\label{cl} \cl \,=\,\la+ \sum_{i=1}^\infty
u_{i+1}\la^{-i}
\ee
where the functions $u_i$ depend on the infinite set of the time
variables $(t_1, t_2, t_3, \ldots)=\{t\}$ and the evolution along
these times is determined by the Lax equations
\be\label{cLax}
\frac{\d\cl}{\d t_i}\,=\,\{\cl^i_+,\cl\}\hspace{0.5cm}i=1, 2,\ldots
\ee
where the functions $\cl^i_+(\{t\},\la)$ are polynomials in $\la$, and, in
complete analogy with the standard KP theory, are defined as a non-negative
parts of the corresponding degrees of the $\cl$-operator:
\be
\cl^i_+\,\equiv \cl^i-\cl^i_-
\ee
In (\ref{cLax}) the Poisson bracket $\{.,.\}$ is the quasiclassical analog
of the commutator; for any functions $F(t_1,\la),\;G(t_1,\la)$
\be
\{F,G\}\,=\,\frac{\d F}{\d\la}\frac{\d G}{\d t_1}\,-\,
\frac{\d F}{\d t_1}\frac{\d G}{\d\la}
\ee
It is useful to introduce the additional operator
\be\label{cm}
\cm\,=\,\sum_{n=1}^\infty nt_n\cl^{n-1}+\sum_{i=1}^\infty h_{i+1}\cl^{-i-1}
\,\equiv\,\sum_{i\in\ZZ}^\infty i\,t_i\,\cl^{i-1}
\ee
which satisfies the equations
\footnote{One should stress that "the negative times" $t_{-i}\,,\;i>0$
are the functions of the independent set $\{t_i\,,\;i>0\}$ which are
determined by the evolution equations (\ref{lm1}). They have
nothing to do with the actual negative times of the Toda lattice hierarchy.}

\be\label{lm1}
\frac{\d\cm}{\d t_i}\,=\,\{\cl^i_+,\cm\}\hspace{0.5cm}i=1, 2,\ldots
\ee
\be\label{lm2}
\{\cl,\cm\}\,=\,1
\ee
Originally, the differential prototype of (\ref{cm}) for KP hierarchy
was introduced in \cite{Orlov} in order to describe the symmetries
of the evolution equations.\\
In \cite{Kri1}, \cite{TT} it was proved that there exists the function
$S(\{t\},\la)$ whose total derivative is given by
\be
dS\,=\,\sum_{i=1}^\infty \cl^i_+dt_i\,+\,\cm d_\la\cl
\ee
and, consequently,
\be\label{ds}
\Big(\frac{\d S}{\d t_i}\Big)_\cl\,=\,\cl^i_+\;;\ \ \ \ \ \
d_\la S\,=\,\cm d_\la\cl
\ee
The function $S$ is a direct quasiclassical analog of the logarithm of the
Baker-Akhiezer function; the solution to (\ref{ds}) can be represented in the
form \cite{TT}
\be\label{S}
S\,=\,\sum_{n=1}^\infty t_n\cl^n-\sum_{j=1}^\infty \frac{1}{j}h_{j+1}\cl^{-j}
\,\equiv\,\sum_{j\in\ZZ} t_j\,\cl^j
\ee

\subsection{Quasiclassical $\tau$-function and $p$-reduction}
The notion of the quasiclassical $\tau$-function can be introduced as
follows. In \cite{TT} it was proved that
\be\label{der3}
\frac{\d h_{i+1}}{\d t_j}\,=\,\frac{\d h_{j+1}}{\d t_i}\,=\,\mbox{Res}\,
\cl^id_\la\cl^j_+\;;\hspace{1cm}i,j\ge 1
\ee
where the residue operation is defined in (\ref{res}), (\ref{res0}).
Therefore, there exists some function whose derivatives w.r.t. $t_i$
coincide with $h_{i+1}$. By definition, the quasiclassical $\tau$-function
is defined by relations
\be\label{ctau}
h_{i+1}\,=\,\frac{\d\log\tau}{\d t_i}\;\,;\hspace{1cm}i\ge 1
\ee
In \cite{TT2} it was shown that $\tau$-function defined above do satisfies
some dispersionless variant of the bilinear Hirota equations, so the
definition (\ref{ctau}) is reasonable.

\bigskip\noindent
Let us consider $p$-reduced quasiclassical KP hierarchy;
this means that for some
natural $p$ the function $\cp\,\equiv\,\cl^p$ is a polynomial in $\la$, i.e.
\be\label{red1}
\cp_{_-}=0
\ee
One can construct the "dual" function
\be \label{cq}
\cq\,=\,\frac{1}{p}\cm\cl^{1-p}\,\equiv\,
\frac{1}{p}\,\sum_{j\in\ZZ}jt_j\cl^{j-p}
\ee
which satisfy the equation
\be\label{pq}
\{\cp,\cq\}=1
\ee
as a corollary of (\ref{lm2}). In \cite{Kri1}, \cite{TT} the particular
case of $p$-reduced hierarchy has been discussed, namely, when
the function $\cq(\{t\},\la)$ is also polynomial in $\la$:
\be\label{red2}
\cq_{_-}\,=\,0
\ee
This constraint restricts the possible solutions of $p$-reduced
quasiclassical KP hierarchy to very specific subset; as Krichever has
shown \cite{Kri2} the $\tau$-function satisfies the infinite set of the
quasiclassical $W$-constraints when (\ref{red2}) holds
\footnote{Equations (\ref{red1}), (\ref{red2}) and (\ref{pq}) are the
analogues of the Douglas equations \cite{Doug} which, in turn, are
equivalent \cite{FKN1} to the $W$-constraints in the KP hierarchy.}.
In particular, the $\tau$-function satisfies the $\L_{-1}$-constraint
\be\label{con1}
\frac{1}{2}\sum_{i=1}^{p-1}i(p-i)t_it_{p-i}+\sum_{i=1}^\infty (p+i)t_{p+i}
\frac{\d\log\tau}{\d t_i}\,=\,0
\ee
(see the proof below).

\subsection{Quasiclassical times and the structure of solutions}
When constraints (\ref{red1}), (\ref{red2}) are satisfied, one can
construct the solutions of the hierarchy as follows \cite{Kri2}.
Note that, evidently,
\be\label{dd}
\mbox{Res}\,\cl^{i-1}d_\la \cl\,=\,\delta_{i,0}
\ee
Multiplying (\ref{cq}) by $\cl^{p-i-1}d_\la\cl$ and taking the residue
with the help (\ref{res}) it is easy to show that
\be\label{ct1}
t_i\,=\,-\,\frac{p}{i(p-i)}\,\mbox{\rm Res}\,\cl^{p-i}d_\la\cq
\hspace{1cm}i\in\ZZ
\ee
From these equations for $i>0$ one can determine (at least, in principle)
the coefficients of $\cq=\sum_{i=0}^\infty q_i\la^i$ and $\cl$ as the functions
of times $t_1, t_2, \ldots$, while the same equations for $i<0$ give then
the parametrization of the "negative times" $t_{-i}=-\frac{1}{i}h_{i+1}$:
\be\label{neg}
t_{-i}\,=\,-\,\frac{1}{i}\,\frac{\d\log\tau}{\d t_i}
\ee
in terms of $t_1, t_2, \ldots$. Consider the simplest situation when
$\cq(\la)$ is a polynomial of the first order. From (\ref{cq}), (\ref{red2})
it is easily seen that such condition is equivalent to switching off
all the times with $i>p+1$: $t_{p+2}=t_{p+3}=\ldots =0$. In this case
\be
\cq(\{t\},\la)\,=\,\frac{p+1}{p}\,t_{p+1}\la\,+\,t_p
\ee
and equations (\ref{ct1}) are reduced to
\be\label{ct2}
t_{i}\,=\,-\,\frac{(p+1)t_{p+1}}{i(p-i)}\,\mbox{Res}\,
\cp^{\frac{p-i}{p}}(\la)d\la\;;\hspace{1cm}i\leq p+1
\ee
Equations (\ref{ct2}) determine the coefficients of the
polynomial $\cp$ as the functions of first $p+1$ times $t_1,t_2,\ldots ,
t_{p+1}$.

It is easy to see that the first time $t_1$ is contained (linearly) only
in $\la$-independent term of $\cp(t,\la)$. Therefore,
\be
\frac{\d\cp}{\d t_1}\,=\,-\frac{p+1}{p}\,t_{p+1}\;;\hspace{1cm}
\frac{\d\cl^i_+}{\d t_1}\,=\,0\;\ \ (i=1, \ldots , p)
\ee
The Lax equations (\ref{cLax}) are reduced now to the form
\be\label{cLax2}
\frac{\d\cp}{\d t_i}\,=\,-\,\frac{\d\,\cp^{i/p}_{_+}}{\d\la}
\cdot\frac{p}{(p+1)t_{p+1}}
\hspace{1cm}i=1, \ldots , p
\ee
The rest of equations (\ref{ct2}) determine the functions
$t_{-i}(t_1,\ldots,t_{p+1})\,,\;\,i\geq 1$. It is possible to find
the explicit time dependence straightforwardly. For example,
using the equation of motion (\ref{cLax2}), one gets
\be\label{ti}
-j\frac{\d t_{-j}}{\d t_i}\,=\,
\mbox{Res}\,{\cal P}^{j/p}d_\la{\cal P}^{i/p}_{_+}\,=\,
\mbox{Res}\,{\cal P}^{i/p}d_\la{\cal P}^{j/p}_{_+}
\ee
where one uses the properties (\ref{res0}) of the residue operation.
In particular, the differentiation of $t_{-1}$ leads to the
simple relation (since $d_\la{\cal P}^{1/p}_{_+}\equiv d\la$)
\be
\frac{\d t_{-1}}{\d t_i}\,=\,-\,\mbox{Res}\,\cp^{i/p}d\la\,=\,
\frac{i(p-i)}{p+1}\,\frac{t_{p-i}}{t_{p+1}}
\ee
After integration of these equations one arrives to relation
\be\label{t-1}
t_{-1}\,=\,-\,\frac{\d\log\tau}{\d t_1}\,=\,\frac{1}{2(p+1)t_{p+1}}\,
\sum_{i=1}^{p-1}i(p-i)t_it_{p-i}
\ee
which is equivalent to $\L_{-1}$-constraint (\ref{con1}) with
$t_{p+2}=\ldots =0$.

\bigskip\noindent
Note, that without loss of generality one can choose
\be\label{tp}
t_{p+1}= \frac{\textstyle p}{\textstyle p+1}
\ee
by the rescaling of lower times
and, therefore, the main equations (\ref{ct2}), (\ref{cLax2}) acquire
the standard form
\footnote{
Note also that the function $\cp$ does not contain the term proportional
to $\la^p$ due to the structure of the $\cl$-operator (\ref{cl}),
hence, $\d\cp/\d t_p=0$.}
\be\label{ct2'}
t_{i}\,=\,-\,\frac{p}{i(p-i)}\,\mbox{Res}\,
\cp^{\frac{p-i}{p}}(\la)d\la\;;\hspace{1cm}i\leq p+1
\ee
\be\label{cLax3}
\frac{\d\cp}{\d t_i}\,=\,-\,\frac{\d\cp^{i/p}_{_+}}{\d\la}
\hspace{1cm}i=1,\ldots , p
\ee

\subsection{Comparison with GKM}
The structure of the quasiclassical hierarchy has a nice interpretation
in the GKM framework. First of all, the prepotential
$V'(\mu)\equiv{\cal P}(\mu)$ of GKM generates the solution of the
quasiclassical KP hierarchy subjected the constraints
\be
{\cal P}_-(\mu)=0\;;\hspace{0.5cm}{\cal Q}(\mu)\,\sim\,\mu
\ee
The easiest way to see this is to note that
the definitions (\ref{s4}) and (\ref{ct2'}) are the same. Moreover,
{\it all} the quasiclassical ingredients are naturally reproduced.
Consider the first basis vector from the set $\{\phi_i^V(\mu)\}$ defined
by (\ref{s9}). Neglecting the exponential prefactor one can easily see
that the object
\be\label{q1}
\Psi(t,\mu)\,=\,\sqrt{p\,\mu^{p-1}}\int e^{-V(x)+x\mu^p}dx
\ee
is a Baker-Akhiezer function of the $p$-reduced quasiclassical KP hierarchy
\footnote{The use of $\mu$ instead of $\wt{\mu}$ should not lead to
confusion.} (recall that the coefficients of $V$ are parametrized by the
quasiclassical times according to (\ref{s4})).
It is evident that $\Psi(t,\mu)$ has the usual asymptotic
\be\label{q2}
\Psi(t,\mu)\stackreb{\mu\rightarrow\infty}{\longrightarrow}
\exp\Big(\sum^{\infty}_{k=1}t_k\mu^k\Big)\left(1+O(\mu ^{-1})\right)
\ee
Using equations of motion for quasiclassical KP hierarchy
\be\label{q3}
\frac{\d V}{\d t_k}\,=\,-\,{\cal P}^{k/p}_+
\ee
(this is the consequence of equations (\ref{cLax3}) or, {\it equivalently}
the corollary of parameterization (\ref{ct2'})) one can easy to show that
the Baker-Akhiezer function (\ref{q1}) satisfies the usual equations of the
$p$-reduced KP hierarchy:
\be\label{q4}
\Big[{\cal P}(\d_{t_1}) + t_1\Big]\,\Psi(t,\mu)\,=\,\mu^p\Psi(t,\mu)\\
{\d\Psi(t,\mu)\over\d t_i}\,=\,
{\cal P}^{k/p}_+(\d_{t_1})\,\Psi(t,\mu)
\ee
where polynomials ${\cal P}^{k/p}_+(\mu)$ are functions of times
$t_1,\,\ldots\, t_p$. Hence, the function (\ref{q1}) gives the explicit example
of exact solution. On the other hand, it
is important that ${\cal P}^{k/p}_+(\mu)$ does not depend on $t_1$ for
$k<p$ and, therefore, in the corresponding equations
(\ref{q4}) we can
treat $\d /\d t_1$ as a formal {\it parameter}, not an operator, i.e.
it is a case of quasiclassical system. Thus,
we see that "the quasiclassical limit" can be naturally treated
in the (pseudo-differential) context of the standard hierarchy
and quasiclassical solutions are
{\it exact} solutions of the full $p$-reduced KP hierarchy restricted on the
"small phase space". The exact Baker-Akhiezer function (\ref{q1}) gives the
explicit solution of quasiclassical evolution equations along first $p$
flows since the standard relation holds:
\be\label{q5}
\Psi(t,\mu)\,=\,\exp\Big(\sum^{p+1}_{k=1}t_k\mu^k\Big)\,
\frac{\tau\Big(t_k-\frac{\textstyle 1}{\textstyle k\mu^k}\Big)}{\tau(t_k)}
\ee
Evaluating the Baker-Akhiezer function (\ref{q1}) by the steepest descent
method, it is possible to find all the derivatives of the $\tau$-function
entering in the r.h.s. of (\ref{q5}). To conclude, the quasiclassical
hierarchy is determined completely by GKM integrals.\\
As a consequence of the above reasoning on can see that
upper $p\times p$ diagonal minor of the matrix
$A_{ij}$ (\ref{s1}) (which appears for the first time in the context of the
equivalent hierarchies) can be written as the second derivative of the
quasiclassical $\tau$-function. Indeed, in the $p$-reduced case the formulas
(\ref{der3}) and (\ref{ctau}) read
\footnote{We denote now the quasiclassical $\tau$-function as $\tau_0$ in
what follows.}
\be\label{q6}
\frac{\d^2\log\tau_0(t)}{\d t_i\d t_j}\,=\,\mbox{Res}\,{\cal P}^{i/p}
d_\la{\cal P}^{j/p}_{_+}\;;\hspace{0.8cm}i,j=1\,,\ldots\,,p
\ee
and the r.h.s. is nothing but (\ref{s1}). Further, the "negative" times
entering to partition function (\ref{s5}) are also represented with the help
of $\tau_0$ due to (\ref{neg}):
\be\label{q7}
kt_{-k}\,=\,-\,\frac{\d \log\tau_0(t)}{\d t_k}\;;
\hspace{0.8cm}k=1\,,\ldots\,,p
\ee
Before returning to the GKM $\tau$-function we need to prove some useful
statement concerning the homogeneity of the quasiclassical $\tau$-function.

\subsection{Homogeneity property}
\begin{lem}
The conditions (\ref{red1}), (\ref{red2}) imply
\be\label{S2}
S_{_-}\,=\,0
\ee
\end{lem}
{\bf Proof.}~~ Recall that $d_\la S=\cm d_\la\cl$. Therefore,
\be
\mbox{Res}\,\cl^i_+ d_\la S\,=\,\mbox{Res}\,\cl^i_+\cm d_\la\cl\,\equiv\,
\,p\,\mbox{Res}\,\cl^i_+\Big(\frac{1}{p}\cm\cl^{1-p}\Big)\cl^{p-1}
d_\la\cl\,\equiv \mbox{Res}\,\cl^i_+\cq d_\la \cp
\ee
Since, by definition, $\cp_{_-}=0$, one gets further
\be
\mbox{Res}\,\cl^i_+\cq d_\la \cp\,\,=\,\mbox{Res}\,(\cl^i_+\cq)_{_-}d_\la
\cp\,=\,\mbox{Res}\,(\cl^i_+\cq_{_-})_{_-}d_\la \cp
\ee
On the other hand,
$\mbox{Res}\,\cl^i_+d_\la S=\mbox{Res}\,\cl^i_+ d_\la S_{_-}$.
Thus, finally,
\be
\mbox{Res}\,\cl^i_+ d_\la S_{_-}\,=\,
\mbox{Res}\,(\cl^i_+\cq_{_-})_{_-}d_\la \cp
\hspace{1cm}\forall\,i=1,2,\ldots
\ee
Therefore, if $\cq_{_-}=0$ then $\mbox{Res}\,\cl^i_+(d_\la S)_{_-}=0$ for
any $i=1,2,\ldots $. The last equality is equivalent to
\be
\mbox{Res}\,\la^i d_\la S_{_-}=0
\ee
and, consequently, (\ref{S2}) holds.\square
\begin{lem}
The constraint (\ref{S2}) is equivalent to homogeneity condition
\be\label{hom}
\sum_{n=1}^\infty t_n\frac{\d t_{-i}}{\d t_n}\,=\,t_{-i}
\ee
\end{lem}
{\bf Proof.}~~ Using the explicit representation of $S$  one gets
\be
\mbox{Res}\,Sd_\la\cl^i_+\,=\,\sum_{n=1}^\infty t_n\mbox{Res}\,
\cl^nd_\la \cl^i_+
-\sum_{j+1}^\infty\frac{1}{j}\,h_{j+1}\mbox{Res}\,\cl^{-j}d_\la \cl^i_+
\ee
But for $j>0$ $\mbox{Res}\,\cl^{-j}d_\la \cl^i_+\,\equiv\,
\mbox{Res}\,\cl^{-j}d_\la (\cl^i-\cl^i_{_-})=
\mbox{Res}\,\cl^{-j}d_\la \cl^i
=i\mbox{Res}\,\cl^{i-j-1}d_\la\cl\,=\,i\delta_{ij}$ due to (\ref{dd})
and, using (\ref{der3}), we have
\be
\mbox{Res}\,Sd_\la\cl^i_+\,=\,\sum_{n=1}^\infty t_n
\frac{\d h_{i+1}}{\d t_n}-h_{i+1}
\ee
or, equivalently
\be
\mbox{Res}\,\cl^i_+d_\la S_{_-}\,=\,h_{i+1}-
\sum_{n=1}^\infty t_n\frac{\d h_{i+1}}{\d t_n}
\ee
and in the case $S_{_-}=0$ one arrives to (\ref{hom})
using the identification $h_{i+1}=-it_{-i}$.\square
Note that in terms of the $\tau$-function (\ref{ctau})
the homogeneity condition (\ref{hom}) has the form
\be\label{hom2}
\sum_{n=1}^\infty t_n\frac{\d\log\tau_0}{\d t_n}\,=\,2\,\log\tau_0
\ee

\section{Polynomial GKM: synthesis}
\subsection{$\L_{-p}$-constraint}\label{pvir}

Here we represent the proof of the Virasoro constraint (\ref{s15}),
(\ref{s16}).

\bigskip\noindent
Let $\phi^{(can)}_i$ be the canonical basis vectors corresponding to
GKM ones, $\phi^V_i$, defined by (\ref{s9}). From the general formula
(\ref{3der}) one gets the expression for the derivative w.r.t. first time
$\wt{T}_1$:
\be\label{vi1}
\frac{\d \tau^V}{\d\wt{T}_1}\,=\,\frac{1}{\Delta(\wt\mu)}\,\left|
\begin{array}{ccc}
\phi^{(can)}_1(\wt{\mu}_1) &\ldots & \phi^{(can)}_1(\wt{\mu}_N)\\
\ldots &\ldots&\ldots \\
\phi^{(can)}_{N-1}(\wt{\mu}_1) &\ldots&\phi^{(can)}_{N-1}(\wt{\mu}_N)\\
\phi^{(can)}_{N+1}(\wt{\mu}_1) &\ldots&\phi^{(can)}_{N+1}(\wt{\mu}_N)
\end{array}\right|\,-\,
\tau^V
\sum_{m=1}^N\wt{\mu}_m
\ee
Now consider the action of the Virasoro generator $\L_{-p}$ in accordance
with (\ref{Vir4}). The operator $A_{-p}(\wt{\mu})$ being expressed via operator
$A(\wt\mu)$ (\ref{s12})
\be
A_{-p}(\wt\mu)\,=\,pA(\wt{\mu})-\sum_{k=1}^{p-1}kt_k\wt{\mu}^{k-p}-
pt_p-p\wt{\mu}
\ee
results to relation
\be\label{vi2}
\frac{1}{p}L_{-p}(\wt{T})\tau^V\,=\,-\,\frac{1}{\Delta(\wt\mu)}\sum_{k=1}^N
A(\wt{\mu}_m)\det\phi^{(can)}_i(\wt{\mu}_j)-\sum_{k=1}^{p-1}k(p-k)\wt{T}_{p-k}
t_k+\tau^V\Big(Nt_p+\sum_{m=1}^N\wt{\mu}_m\Big)
\ee
where the KP times $\wt{T}_k$ are expressed through the Miwa variables
$\wt{\mu}_i$ due to (\ref{gr10}). In order to prove the analog of constraint
(\ref{r12}) one should calculate the action of the operator $A(\wt{\mu})$ on
the canonical basis vectors starting from (\ref{s11}). The vectors (\ref{s9})
are not of canonical form; let
\be\label{vi0}
\phi^V_i(\wt{\mu})\,=\,\wt{\mu}^{i-1}+\al_i\,\wt{\mu}^{i-2}+\ldots\;;
\hspace{0.7cm}i=1, 2,\,\ldots
\ee
(see the exact expression for $\al_i$ below). Now $\phi^{(can)}_i=\phi^V_i-
\al_i\phi^V_{i-1}+\ldots$ and, therefore,
\be\label{vi3}
A(\wt{\mu})\phi^{(can)}_i\,=\,\phi^{(can)}_{i+1}+(\al_{i+1}-
\al_i)\phi^{(can)}_i+\ldots\;;\hspace{1cm}\al_1=0
\ee
It is evident that
\be\label{vi4}
\frac{1}{\Delta(\wt\mu)}\sum_{k=1}^N
A(\wt{\mu}_m)\det\phi^{(can)}_i(\wt{\mu}_j)\,=\,
\frac{1}{\Delta(\wt\mu)}\,\left|
\begin{array}{ccc}
\phi^{(can)}_1(\wt{\mu}_1) &\ldots & \phi^{(can)}_1(\wt{\mu}_N)\\
\ldots &\ldots&\ldots \\
\phi^{(can)}_{N-1}(\wt{\mu}_1) &\ldots&\phi^{(can)}_{N-1}(\wt{\mu}_N)\\
\phi^{(can)}_{N+1}(\wt{\mu}_1) &\ldots&\phi^{(can)}_{N+1}(\wt{\mu}_N)
\end{array}\right|\,+\,\al_{N+1}\,\tau^V
\ee
and, after substitution of (\ref{vi4}), (\ref{vi1}) to (\ref{vi2}), one
arrives to relation
\be\label{vi5}
\Big(\frac{1}{p}\L_{-p}+\frac{\d}{\d \wt{T}_1}\Big)\tau^V\,=\,-\,
\sum_{k=1}^{p-1}k(p-k)\wt{T}_{p-k}t_k\,+(Nt_p-\al_{_{N+1}})\tau^V
\ee
The last step is to calculate the coefficients $\al_i$ in the expansion
(\ref{vi0}). Recall, that the vectors $\phi^V_i$ (\ref{s9}) are related with
the original ones (\ref{gr0}) as follows (taking into account the relation
(\ref{s3})):
\be\label{vi6}
\phi_i^V(\wt\mu)\,=\,\sqrt{\frac{\displaystyle p\wt{\mu}^{p-1}}{V''(\mu)}}\;
e^{\sum_{k=-\infty}^{-1}t_k\wt{\mu}^k}\,\Phi^V_i(\mu)\,=\\ =\,
\sqrt{\frac{\displaystyle p\wt{\mu}^{p-1}}{V''(\mu)}}\;
e^{t_{-1}\wt{\mu}^{-1}}\,\mu^{i-1}\Big(1+O(\mu^{-2})\Big)
\ee
- see the asymptotics (\ref{v2}). Using the residue technique it is easy to
find that
\be
\frac{p\wt{\mu}^{p-1}}{V''(\mu)}\,=\,\sum_{i=-\infty}^{1}\frac{i(p+i)}{p+1}\,
\frac{t_{p+i}}{t_{p+1}}\,\wt{\mu}^{i-1}\,=\,1+O(\wt{\mu}^{-2})
\ee
i.e. this term does not contribute to $\al_i$. Therefore, from (\ref{vi6})
and (\ref{s3'})
\be
\phi^V_i(\wt{\mu})\,=\,\wt{\mu}^{i-1}+\Big(t_{-1}+(i-1)t_p\Big)\wt{\mu}^{i-2}
+\ldots
\ee
Hence, $\al_i=(i-1)t_p+t_{-1}$ and, consequently,
\be
Nt_p-\al_{_{N+1}}\,=\,-t_{-1}
\ee
where $t_{-1}$ is just the corresponding derivative of the quasiclassical
$\tau$-function defined by (\ref{t-1}); using the convention (\ref{tp})
it reads now
\be\label{qVir}
t_{-1}\,=\,\frac{1}{2p}\sum_{k=1}^{p-1}k(p-k)t_kt_{p-k}
\ee
Hence, the equation (\ref{vi4}) acquires the form
\be\label{vi7}
\frac{1}{2p}\sum_{k=1}^{p-1}k(p-k)(\wt{T}_k\!+\!t_k)(\wt{T}_{p-k}\!+\!t_{p-k})
+\frac{1}{p}\sum_{k=1}^\infty(k\!+\!p)(\wt{T}_{k+p}\!+\!t_{p+k})
\frac{\d\log\tau^V}{\d \wt{T}_k}\,=\,0
\ee
and $\L_{-p}$-constraint (\ref{s15}), (\ref{s16}) is proved.

\subsection{Complete description of time dependence}\label{ptimes}

Let us bring all essential facts together.
\begin{itemize}
\item[(i)]{We transformed the original
matrix integral (\ref{gr}) in the terms of new times (\ref{gr12}) as follows:
\be\label{z1}
Z^V[T(\wt{T})]\,\equiv\,\frac{\Delta(\wt\mu)}{\Delta(\mu)}\,
\prod_{i}\left(\frac{V''(\mu_i)}{p\,\wt\mu_i^{p-1}}\right)^{1/2}\wt Z^V[\wt T]
\,=\,e^{-\frac{1}{2}\,A_{ij}\wt{T}_i \wt{T}_j}\,\wt Z^V[\wt T]
\ee
}
\item[(ii)]{ The matrix $||A_{ij}||$ depends on the quasiclassical times
$\{t_i\}$ related with the coefficients of the polynomial ${\cal P}(\la)
\equiv V'(\la)$ by the formula (\ref{ct2'}) and is
compactly written in the form
\be\label{z2}
A_{ij}(t)\,=\,\mbox{Res}\,{\cal P}^{i/p}(\la)d_\la{\cal P}^{j/p}_{_+} (\la)
\ee
Moreover, for $1\leq i,j\leq p$
\be\label{z'}
A_{ij}(t)\,=\,\frac{\d^2\log\tau_0(t)}{\d t_i\d t_j}
\ee
and $\tau_0$ is a $\tau$-function of the $p$-reduced quasiclassical
KP hierarchy.
}
\item[(iii)]{The "preliminary" $p$-reduced $\tau$-function of GKM is a
solution of the equivalent hierarchy; it is represented as
\be\label{z5}
\wt Z^V[\wt T]\,=\,\frac{\Delta(\wt{\mu}^p)}{\Delta(\wt\mu)}
\prod_i(p\,\wt\mu^{p-1}_i)^{1/2}\;\;
e^{\raise2pt\hbox{$\scriptstyle \!\Tr\,[V(M)-MV'(M)]$}}
\int\,e^{\raise2pt\hbox{$\scriptstyle \!\Tr\,[-V(X)+XV'(M)]$}}\;dX
\ee
where $\wt{\mu}^p\equiv V'(\mu)$.
Note that the r.h.s. of (\ref{z5}) is expressed in terms of the
matrix $\wt M$. In order to stress such dependence we denote the
corresponding terms as $M|_{_{\wt M}}$ etc. below.
}
\item[(iv)]{
The time derivatives of the partition function (\ref{z1}) can be also
written as a matrix integrals
\be\label{z4}
\frac{\d}{\d \wt T_k}\log Z^V[T(\wt T)]\,=\,
\<\mbox{Tr}\,[{\cal P}^{k/p}(X)]_{_+}
-\mbox{Tr}\,[{\cal P}^{k/p}(M)]_{_+}\> \hspace{1cm}1\leq k\leq p
\ee
}
\end{itemize}
Let us differentiate $\wt Z^V$ w.r.t. quasiclassical times $t_k$ keeping
$V'(M)=\wt M^p$ fixed. It is clear that
\be
\frac{\d}{\d t_k}\mbox{Tr}\,V(X)\,=\,-\,\mbox{Tr}\,[{\cal P}^{k/p}(X)]_{_+}
\ee
due to quasiclassical evolution equations. To calculate the derivative
of $\{V(M)-MV'(M)\}|_{_{\wt M}}$ one should take into account that,
besides of the coefficient of $V$, the elements of the matrix $M$ are also
dependent on times $\{t_k\}$ as a functions of $\wt M$. Therefore,
\be
\frac{\d}{\d t_k}\mbox{Tr}\,\left[\,V(M)|_{_{\wt M}}-M|_{_{\wt M}}
\,{\wt M}^p\,\right]\,\equiv\\ \equiv\,
\sum_{i=1}^{p+1}\frac{1}{i}\frac{\d v_i(t)}{\d t_k}\mbox{Tr}\,M^k\,+\,
\mbox{Tr}\,\left[\frac{\d V(M)}{\d M}\,\frac{\d
M}{\d t_k} \,-\,\frac{\d M}{\d t_k}\,\wt{M}^p\right]_{\wt M}
\ee
The second term in this expression is vanished identically, thus,
\be
\frac{\d}{\d t_k}\mbox{Tr}\,[V(M)-M{\wt M}^p)]\,=\,-\,\mbox{Tr}\,\left[
{\cal P}_{_+}^{k/p}(M)|_{_{\wt M}}\right]
\ee
and, finally,
\be
\frac{\d}{\d t_k}\log\,\wt Z^V[\wt T]\,=\,
\<\mbox{Tr}\,[{\cal P}^{k/p}(X)]_{_+}
-\mbox{Tr}\,[{\cal P}^{k/p}(M)]_{_+}|_{_{\wt M}}\> \hspace{1cm}1\leq k\leq p
\ee
Therefore, comparing the last relation with (\ref{z4}) it follows that
\be\label{t4}
\frac{\d}{\d T_k}\log Z^V[T(\wt T)]\,=\,\frac{\d}{\d t_k}\log\wt Z^V[\wt T]
\ee
From (\ref{z1}) and (\ref{t4}) one obtains
\be\label{t5}
\left(\frac{\d}{\d \wt{T}_k}-\frac{\d}{\d t_k}\right)\log Z^V[\wt{T},t]\,=\,
\sum_{i=1}^\infty A_{ki}\wt{T}_i\hspace{1cm}1\leq k\leq p
\ee
Introduce the $\tau$-function $\tau^V$ in agreement with (\ref{s5}),
(\ref{s7}):
\be\label{z6}
\wt Z^V[\wt T,t]\,=\,
e\raise9pt\hbox{${\scriptscriptstyle\sum_{i=1}^\infty} {\scriptstyle
it_{-i}\wt{T}_i}$}\,\tau^V[\wt{T},t]
\ee
where negative times $t_{-i}\,,\;\,i=1, 2, \ldots$ satisfy the equations
\be\label{z7}
-\,i\frac{\d t_{-i}}{\d t_k}\,=\,A_{ki}\;;
\hspace{0.8cm}k\,=\,1,\,\ldots\,,p
\ee
in accordance with (\ref{ti}).
Substitution of $\tau^V$ to (\ref{t5}) results to equation
\be\label{z8}
\left(\frac{\d}{\d \wt{T}_k}-\frac{\d}{\d t_k}\right)\log\tau^V[\wt T,t]\,=\,
-\,kt_{-k}\,+\sum_{i=1}^\infty\Big(A_{ki}+i\frac{\d t_{-i}}{\d t_k}\Big)
\wt{T}_i\,=\\ =\,-\,kt_{-k}\,\equiv\,\frac{\d\log\tau_0(t)}{\d t_k}
\ee
due to relations (\ref{z7}). Thus, the partition function
\be\label{z9}
{\cal Z}^V[\wt{T},t]\,\equiv\,\tau^V[\wt{T},t]\,\tau_0(t)
\ee
depends only on the sum $\wt{T}_k+t_k$:
\be\label{z10}
\frac{\d{\cal Z}^V}{\d\wt{T}_k}\,=\,
\frac{\d{\cal Z}^V}{\d t_k}
\ee
Let us find the relation between partition functions $Z^V$ and ${\cal Z}^V$.
Due to (\ref{z1}), (\ref{z6}) and (\ref{z9}) these functions are proportional
up to the exponential with
\be\label{z11}
-\frac{1}{2}\,A_{ij}\wt{T}_i \wt{T}_j+ it_{-i}\wt{T}_i-\log\tau_0(t)
\ee
Due to homogeneity relation (\ref{hom}) the expression (\ref{z11})
can be written as
\be
-\frac{1}{2}\sum_{ij}A_{ij}(t)\,(\wt{T}_i\!+\!t_i)(\wt{T}_j\!+\!t_j)
\ee
hence,
\be\label{z12}
Z^V[T(\wt{T})]\,=\,{\cal Z}^V(\wt{T}+t)\exp\left\{
-\frac{1}{2}\sum_{ij}A_{ij}(t)
(\wt{T}_i\!+\!t_i)(\wt{T}_j\!+\!t_j)\right\}
\ee
This formula gives the complete description of the polynomial GKM w.r.t
"quantum" $(\{\wt{T}_k\})$ and quasiclassical $(\{t_k\})$ times.

\section{Acknowledgments}
The author is deeply indebted to his coauthors
A.Marshakov, A.Mironov and A.Morozov for stimulating discussions.
This work is partly supported by grant RFBR 96-02-19085 and by
grant 96-15-96455 for Support of Scientific Schools.

\end{document}